\documentclass[preprint,amsmath,amssymb,superscriptaddress,longbibliography,aps,floatfix,footinbib]{revtex4-2}
\usepackage[]{graphicx}
\usepackage{tabularx}
\usepackage[usenames,dvipsnames]{color}
\usepackage{soul}
\usepackage{bm}
\usepackage{amsmath}
\usepackage{lineno}
\usepackage{gensymb}

\usepackage{amsfonts}
\usepackage{mathrsfs}
\usepackage{graphicx}
\usepackage{dcolumn}
\usepackage{bm}
\usepackage{color}

\usepackage[colorlinks,bookmarks=false,citecolor=blue,linkcolor=red,urlcolor=blue]{hyperref}
\usepackage{multirow}
\usepackage{physics}
\usepackage{siunitx}
\DeclareSIUnit{\barpressure}{bar}
\DeclareSIUnit\angstrom{\protect \text {Å}}

\begin{document}

\title{Robust spin splitting and fermiology in the layered altermagnet CoNb$_4$Se$_8$}

\author{Alessandro De Vita***}\email{devita@fhi-berlin.mpg.de}
\affiliation{Institut für Physik und Astronomie, Technische Universität Berlin, Straße des 17 Juni 135, 10623 Berlin, Germany}
\affiliation{Fritz Haber Institut der Max Planck Gesellschaft, Faradayweg 4--6, 14195 Berlin, Germany}

\author{Chiara Bigi***}\email{chiara.bigi@synchrotron-soleil.fr}
\affiliation{Synchrotron SOLEIL, F-91190 Saint-Aubin, France}

\author{Davide Romanin***}
\affiliation{Université Paris-Saclay, CNRS, Centre de Nanosciences et de Nanotechnologies, 91120, Palaiseau, Paris, France}

\author{Matthew D. Watson}
\affiliation{Diamond Light Source Ltd, Harwell Science and Innovation Campus, Didcot, OX110DE, United Kingdom}

\author{Vincent Polewczyk}
\affiliation{Universit\'e Paris-Saclay, UVSQ, CNRS, GEMaC, 78000, Versailles, France}

\author{Marta Zonno}
\affiliation{Synchrotron SOLEIL, F-91190 Saint-Aubin, France}

\author{Fran\c cois Bertran}
\affiliation{Synchrotron SOLEIL, F-91190 Saint-Aubin, France}

\author{Pietro M. Forcella}
\affiliation{Université Paris-Saclay, CNRS, Centre de Nanosciences et de Nanotechnologies, 91120, Palaiseau, Paris, France}

\author{My Bang Petersen}
\affiliation{Department of Physics and Astronomy, Interdisciplinary Nanoscience Center, Aarhus University, 8000 Aarhus C, Denmark}

\author{Federico Motti}
\affiliation{CNR-Istituto Officina dei Materiali (IOM), Unità di Trieste, Strada Statale 14, km 163.5, 34149 Basovizza (TS), Italy\looseness=-1}%

\author{Giovanni Vinai}
\affiliation{CNR-Istituto Officina dei Materiali (IOM), Unità di Trieste, Strada Statale 14, km 163.5, 34149 Basovizza (TS), Italy\looseness=-1}%

\author{Manuel Tuniz}
\affiliation{Dipartimento di Fisica, Università degli Studi di Trieste, 34127, Trieste, Italy}

\author{Federico Cilento}
\affiliation{Elettra - Sincrotrone Trieste S.C.p.A., Strada Statale 14, km 163.5, Trieste, Italy}

\author{Mario Cuoco}
\affiliation{CNR-SPIN, c/o Universit\'a di Salerno, IT-84084 Fisciano (SA), Italy}

\author{Brian M. Andersen}
\affiliation{Niels Bohr Institute, University of Copenhagen, 2100 Copenhagen, Denmark}

\author{Andreas Kreisel}
\affiliation{Niels Bohr Institute, University of Copenhagen, 2100 Copenhagen, Denmark}

\author{Luciano Jacopo D'Onofrio}
\affiliation{CNR-SPIN, c/o Universit\'a di Salerno, IT-84084 Fisciano (SA), Italy}

\author{Oliver J. Clark}
\affiliation{School of Physics and Astronomy, Monash University, Clayton, Victoria 3800, Australia}

\author{Mark T. Edmonds}
\affiliation{School of Physics and Astronomy, Monash University, Clayton, Victoria 3800, Australia}

\author{Christopher Candelora}
\affiliation{Department of Physics, Boston College, Chestnut Hill, MA 02467, USA}

\author{Muxian Xu}
\affiliation{Department of Physics, Boston College, Chestnut Hill, MA 02467, USA}

\author{Siyu Cheng}
\affiliation{Department of Physics, Boston College, Chestnut Hill, MA 02467, USA}

\author{Alexander LaFleur}
\affiliation{Department of Physics, Boston College, Chestnut Hill, MA 02467, USA}

\author{Tommaso Antonelli}
\affiliation{ETH Zürich, HPF E 19 Otto-Stern-Weg 1, 8093 Zürich, Switzerland}

\author{Giorgio Sangiovanni}
\affiliation{Institute for Theoretical Physics and Astrophysics, University of W\"urzburg, D-97074 W\"urzburg, Germany}

\author{Lorenzo Del Re}
\affiliation{Institute for Theoretical Physics and Astrophysics, University of W\"urzburg, D-97074 W\"urzburg, Germany}

\author{Ivana Vobornik}
\affiliation{CNR-Istituto Officina dei Materiali (IOM), Unità di Trieste, Strada Statale 14, km 163.5, 34149 Basovizza (TS), Italy\looseness=-1}%

\author{Jun Fujii}
\affiliation{CNR-Istituto Officina dei Materiali (IOM), Unità di Trieste, Strada Statale 14, km 163.5, 34149 Basovizza (TS), Italy\looseness=-1}%

\author{Fabio Miletto Granozio}
\affiliation{CNR-SPIN, c/o Complesso di Monte S. Angelo, IT-80126 Napoli, Italy}

\author{Alessia Sambri}
\affiliation{CNR-SPIN, c/o Complesso di Monte S. Angelo, IT-80126 Napoli, Italy}

\author{Emiliano Di Gennaro}
\affiliation{Physics Department, University of Napoli “Federico II”, Via Cinthia, 21, Napoli 80126, Italy}

\author{Jeppe B. Jacobsen} 
\affiliation{Nanoscience Center, Niels Bohr Institute, University of Copenhagen, 2100 Copenhagen, Denmark}

\author{Henrik Jacobsen} 
\affiliation{European Spallation Source ERIC - Data Management and Software Center, 2800 Kgs.~Lyngby, Denmark}

\author{Iulia Cojocariu}
\affiliation{Elettra - Sincrotrone Trieste S.C.p.A., Strada Statale 14, km 163.5, Trieste, Italy}
\affiliation{Dipartimento di Fisica, Università degli Studi di Trieste, 34127, Trieste, Italy}

\author{Marcin Szpytma}
\affiliation{Elettra - Sincrotrone Trieste S.C.p.A., Strada Statale 14, km 163.5, Trieste, Italy}
\affiliation{Faculty of Physics and Applied Computer Science, AGH University of Krakow, Krakow, Poland}

\author{Andrea Locatelli}
\affiliation{Elettra - Sincrotrone Trieste S.C.p.A., Strada Statale 14, km 163.5, Trieste, Italy}

\author{Tevfik Mentes}
\affiliation{Elettra - Sincrotrone Trieste S.C.p.A., Strada Statale 14, km 163.5, Trieste, Italy}

\author{Won Seop Song}
\affiliation{Department of Physics, Pohang University of Science and Technology, 77 Cheongam-ro, Nam-gu, Pohang-si, Gyeongsangbuk-do, Korea 37673}
\affiliation{Center for Artificial Low-Dimensional Electronic Systems, Institute for Basic Science, Campus 79, Jigok-ro 127beon-gil, Namgu, Pohang, Korea, (37673)}

\author{Jun Sung Kim}
\affiliation{Department of Physics, Pohang University of Science and Technology, 77 Cheongam-ro, Nam-gu, Pohang-si, Gyeongsangbuk-do, Korea 37673}
\affiliation{Center for Artificial Low-Dimensional Electronic Systems, Institute for Basic Science, Campus 79, Jigok-ro 127beon-gil, Namgu, Pohang, Korea, (37673)}

\author{Matthieu Jamet}
\affiliation{Univ. Grenoble Alpes, CEA, CNRS, IRIG-SPINTEC, 38000 Grenoble, France}

\author{Jean-François Jacquot}
\affiliation{Univ. Grenoble Alpes, CEA, CNRS, IRIG-SPINTEC, 38000 Grenoble, France}

\author{Pasquale Orgiani}
\affiliation{CNR-Istituto Officina dei Materiali (IOM), Unità di Trieste, Strada Statale 14, km 163.5, 34149 Basovizza (TS), Italy\looseness=-1}%

\author{Ralph Ernstorfer}
\affiliation{Institut für Physik und Astronomie, Technische Universität Berlin, Straße des 17 Juni 135, 10623 Berlin, Germany}
\affiliation{Fritz Haber Institut der Max Planck Gesellschaft, Faradayweg 4--6, 14195 Berlin, Germany}

\author{Ilija Zeljkovic}\email{ilija.zeljkovic@bc.edu}
\affiliation{Department of Physics, Boston College, Chestnut Hill, MA 02467, USA}

\author{Younghun Hwang}\email{younghh@uc.ac.kr}
\affiliation{Electricity and Electronics and Semiconductor Applications, Ulsan College, Ulsan 44610, Republic of Korea}

\author{Matteo Calandra}\email{m.calandrabuonaura@unitn.it}
\affiliation{Department of Physics, University of Trento, Via Sommarive 14, Povo 38123, Italy}

\author{Jill A. Miwa}\email{miwa@phys.au.dk}
\affiliation{Department of Physics and Astronomy, Interdisciplinary Nanoscience Center, Aarhus University, 8000 Aarhus C, Denmark}

\author{Federico Mazzola}\email{federico.mazzola@unipd.it}
\affiliation{Department of Physics and Astronomy ‘Galileo Galilei’, University of Padova, Padova, Italy}
\affiliation{CNR-Istituto di Struttura della Materia (CNR-ISM), Strada Statale 14, km 163.5, 34149 Trieste, Italy}
\affiliation{CNR-SPIN, c/o Complesso di Monte S. Angelo, IT-80126 Napoli, Italy}

\begin{abstract}
\bf{Altermagnetism defies conventional classifications of collinear magnetic phases, combining compensated magnetization with momentum-dependent, nonrelativistic spin splitting \cite{Smejkal2022_1, Smejkal2022_2, Bai2024, Maier2023, Jungwirth2024,Song2025}. While altermagnetic band splitting has been established in three-dimensional compounds such as MnTe and CrSb \cite{Krempasky2024, Reimers_2024}, layered altermagnets remain far less explored spectroscopically \cite{Regmi2024,Dale_2024,Sakhya2025}. Here, combining termination-resolved micro-ARPES, fine photon-energy mapping, spin-ARPES and density functional theory, we map the three-dimensional fermiology of the layered \textit{g}-wave altermagnet CoNb$_4$Se$_8$, obtained by ordered Co intercalation of 2H-NbSe$_2$. We resolve the spin-split branches of the altermagnetic pairs directly in the raw photoemission spectra and, with spin-ARPES at these experimentally identified momenta, show the reversal of the spin polarization at opposite momenta and between two regions of the sample, consistent with opposite time-reversal domains, and its absence along the symmetry-protected nodal $\Gamma$--K direction. We further report a weak low-temperature anomaly near $T_0\approx\SIrange{50}{60}{\K}$, detected by magnetometry, temperature-dependent micro-ARPES and ultrafast reflectivity, whose microscopic origin remains open. These results provide a momentum-resolved benchmark of the electronic structure of a layered altermagnet.}
\end{abstract}

\maketitle
\noindent
*** These authors contributed equally.

Altermagnetism, recognized as the third elementary type in the classification of non-relativistic spin-group symmetries, is defined by the coexistence of zero-net magnetization and the lifting of Kramers degeneracies, which lead to distinctive transport phenomena \cite{Mazin2021,Mazin2022, Smejkal2022_2,Reichlova2024}. These degeneracies are shaped by opposite spin sublattices, interconnected through rotational symmetry in real space \cite{Smejkal2022_1,Yuan2023,Krempasky2024}. Distinct from both ferromagnetism and antiferromagnetism, this newly identified phase has sparked considerable interest, presenting new challenges for the development of spin-based technologies with potentially groundbreaking properties. The absence of net magnetization offers an advantage over conventional ferromagnetic technologies by mitigating stray magnetic fields. Simultaneously, the lifting of spin degeneracies, coupled with momentum-dependent spin-locking, facilitates efficient spin-filtering. Moreover, contrary to traditional collinear antiferromagnets, altermagnets intrinsically break time-reversal symmetry, giving rise to phenomena such as the anomalous Hall effect, highlighting their potential for novel electronic applications \cite{Fedchenko2024, Smejkal2020, Feng2022, Betancourt2023}.

This rapidly evolving field presents compelling opportunities to extend this phenomenology to two dimensional (2D) material systems \cite{Zeng2024, Mazin2024}. 2D and thin-film altermagnets, along with their layered analogues, are particularly intriguing, envisioned as altermagnetic counterparts to graphene and graphite. \cite{Regmi2024}. While significant progress has been made, the high symmetry of most 2D antiferromagnets prevents the stabilization of the altermagnetic phase \cite{Autieri2022,Basnet2022,Linhart2023} and scalable layer-by-layer synthesis methods remain scarce \cite{Zhou2024}, highlighting the need for innovative approaches. Building on the prediction of altermagnetism through site-ordered cobalt intercalation of the superconducting polymorph 2H-NbSe$_2$ \cite{Regmi2024}, we employ a combined theoretical and experimental approach to demonstrate its altermagnetic band splitting and spin texture.

The altermagnetic ground state of CoNb$_4$Se$_8$ was predicted, and its A-type order determined by neutron diffraction, by Regmi \textit{et al.} \cite{Regmi2024}. Dale \textit{et al.} reported spin-resolved photoemission evidence for nonrelativistic spin splitting, which they distinguish from a relativistic splitting near K by its nodal structure and temperature dependence \cite{Dale_2024}, and Sakhya \textit{et al.} spin-integrated spectra consistent with it \cite{Sakhya2025}. In both works, however, the spin-split branches are not resolved separately in the spectra and are identified through the comparison with DFT, so that the spin signal cannot be assigned to specific bulk bands from the data alone. Here we resolve the two branches of the altermagnetic pairs directly in the raw spin-integrated data (Fig.~\ref{fig2}c--g), measure their spin polarization at these experimentally identified momenta, show that it is finite and reversing along $\Gamma$--M, where any relativistic $\sigma_z$ is forbidden by symmetry, and compatible with zero along the nodal $\Gamma$--K direction, where it vanishes in the bulk, and show its reversal between two regions of the sample, consistent with time-reversal domains. Finally, we report a weak low-temperature anomaly near $T_0\approx\SIrange{50}{60}{\K}$, observed in magnetometry, in temperature-dependent micro-ARPES as a renormalization of the bandwidth, and in ultrafast reflectivity as a sign change of $\Delta R/R$. Its microscopic origin remains open.

\begin{figure}
\includegraphics[width=1.0\columnwidth]{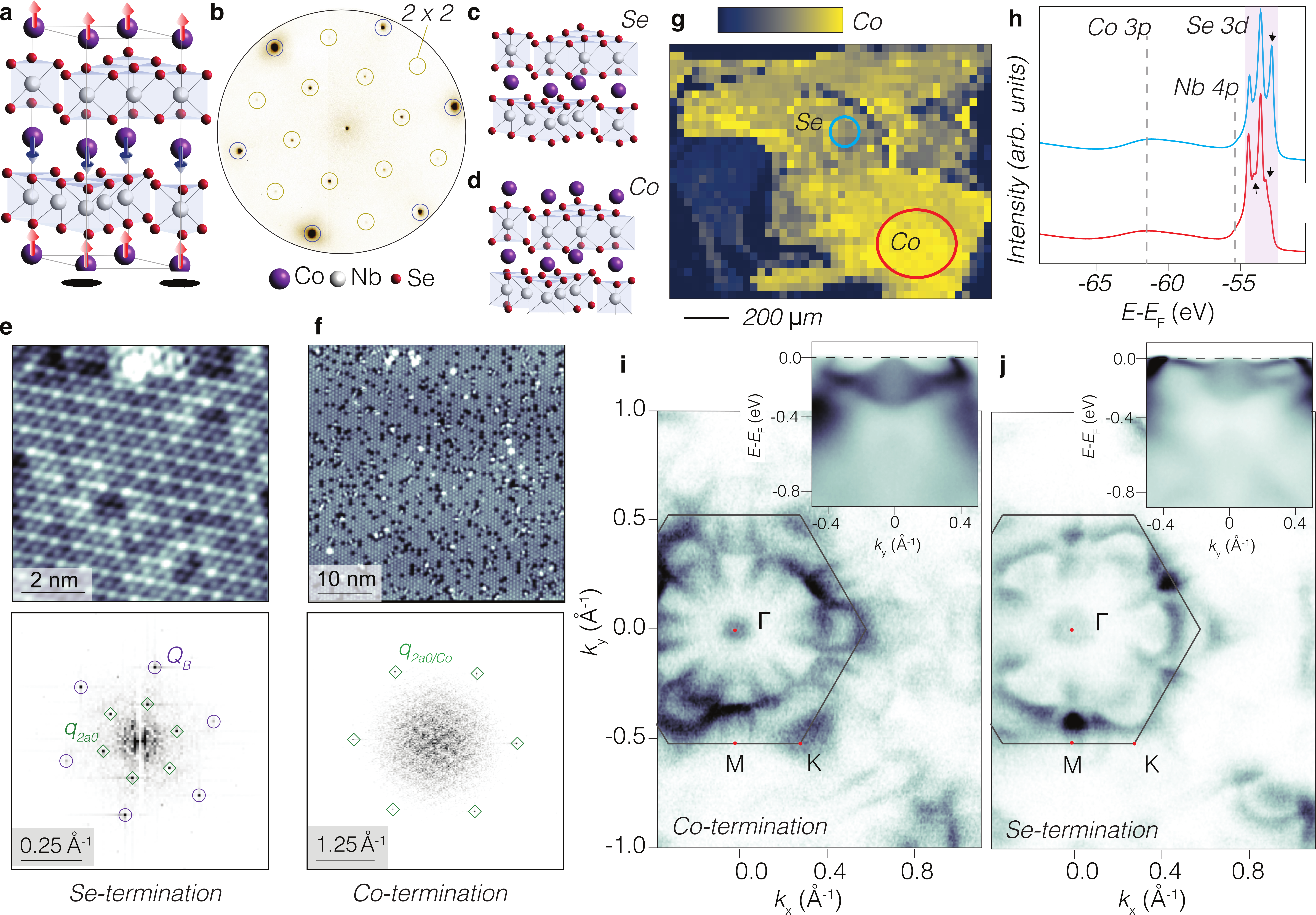}
  \caption{\textbf{Surface-dependent crystalline and electronic structure.} \textbf{a.} Schematic of CoNb$_{4}$Se$_{8}$ formed by 2H-NbSe$_2$ planes linked by Co atoms in a $2\times2$ reconstruction. \textbf{b.} LEED data (\SI{25}{\eV}, $\sim\SI{120}{\K}$) showing the $2\times2$ reconstruction: the orange circles indicate the $2\times2$ features, while the blue circles are the primitive ones. \textbf{c--d.} Schematics showing the possible surface terminations (Se and Co) after cleaving. \textbf{e.} Se-termination (characterized by triangular tiling) and \textbf{f.} Co-termination, measured by STM, and their corresponding Fourier transforms. \textbf{g.} micro-ARPES spatial map acquired by measuring the Se 3\textit{d} and Co 3\textit{p} core levels; yellow corresponds to  Co-terminated regions. \textbf{h.} Core levels acquired from the two different surface terminations (curve colors matching the circle colors in \textbf{g}); the highlighted spectral features are attributed to surface replicas for the Co-termination. \textbf{i--j.} Corresponding Fermi surfaces and (\textit{E}, \textit{k}) dispersions, collected at \SI{25}{\K} with \textit{p}-polarized photons at \SI{47.5}{\eV} (black indicates high intensity in the ARPES scale).}
  \label{fig1}
\end{figure}

Single crystals of CoNb$_4$Se$_8$ were synthesized using the chemical vapour transport (CVT) method; all the data presented in the following have been acquired on samples from a single growth batch. Their composition and phases were monitored by X-ray diffraction (XRD), magnetic susceptibility, specific heat capacity, and X-ray photoelectron spectroscopy (XPS) measurements (Supporting Information Figs.~S1--S4 \cite{SupplementalMaterial}). The crystal structure depicted in Fig.~\ref{fig1}a reveals Co-intercalated NbSe$_2$ planes, which doubles the unit cell of the system, giving a $2\times2$ modulation. Such a $2\times2$ modulation is clearly visible at room temperature, as confirmed by XRD data (Supporting Information Fig.~S1), and persists in the altermagnetic phase, as revealed by low-energy electron diffraction (LEED) (Fig.~\ref{fig1}b, and Supporting Information Fig.~S1). This confirms that the unit cell doubling is an inherent feature of the crystallographic structure. Both our bulk (XRD) and surface (LEED) diffraction probes show exclusively the $2\times2$ superstructure on top of the primitive $1\times1$ NbSe$_2$ pattern, with no detectable $\sqrt{3}\times\sqrt{3}\,R30^\circ$ reflections, \textit{i.e.}, the structural fingerprint of a CoNb$_{3}$Se$_6$-type sublattice \cite{mandujano2025jacs,mandujano2025prm,Parkin1983,VanLaar1971}. Any Co$_{1/3}$-type ordered intergrowth must therefore lie below the detection limit of these probes and be structurally disordered. We can therefore rule it out as the origin of the sharp dispersive features observed throughout our (micro-)ARPES dataset, which instead arise unambiguously from a well-ordered $2\times2$ phase consistent with stoichiometric CoNb$_4$Se$_8$. Spectroscopic measurements further reveal $2\times2$ backfolded electronic features, whose intensity diminishes progressively from \SI{150}{\K} and vanishes at higher temperatures (Supporting Information Figs.~S8-S9). This temperature dependence may originate from surface-related effects or, more trivially, from thermal broadening reducing the spectral resolution. Nevertheless, these changes do not affect the central conclusions of this study, which are drawn from bulk-sensitive observations.

Similar to other intercalated transition metal dichalcogenides (TMDs) \cite{Edwards2023},  CoNb$_4$Se$_8$ cleaves to reveal two distinct surface terminations, each displaying markedly different electronic properties. Figs.~\ref{fig1}c--\ref{fig1}d present schematics of these two terminations, which were experimentally distinguished using scanning tunneling microscopy (STM). The Se-terminated surface, which appears to be the cleanest, exhibits the characteristic triangular tiling of the 2H-NbSe$_2$ polymorph (Fig.~\ref{fig1}e).  In contrast, the Co-terminated surface is typically more disordered yet remains clearly distinguishable from the Se-terminated surface (Fig.~\ref{fig1}f). Additional STM data are provided in Supporting Information Fig.~S3.

We used angle-resolved photoelectron spectroscopy with few-micron spatial resolution (micro-ARPES) to validate these observations. Fermi surfaces for each surface termination were acquired by mapping the surface composition through Se 3\textit{d} and Co 3\textit{p} core level measurements (Fig.~\ref{fig1}g). Spatially resolved maps indicate that each homogeneously terminated region spans approximately \SI{20}{\micro\metre} laterally. According to our data in Fig.~\ref{fig1}h, the Se-termination (blue curve) exhibits surface replicas of the 3\textit{d} core levels, whereas Co-terminated regions (red curve) show distinct spectral features, including an additional peak at lower energies possibly formed by a different coordination of Co-Se atoms. It is important to note that the cleaving process can introduce random surface vacancies, which are readily identifiable in the core level spectra and may locally degrade the ARPES signal. In any case, the altermagnetic phenomena discussed here are of bulk origin and remain unaffected by either surface terminations or surface disorder.

The Fermi surface maps, along with the energy and momentum (\textit{E}, \textit{k}) dispersions along the $\Gamma-\mathrm{M}$ direction (Figs.~\ref{fig1}i--\ref{fig1}j; black indicates the presence of electrons), reveal a complex and rich electronic structure. For both terminations, the intensity displays a pronounced threefold symmetry, accompanied by multiple petals and back-folded features. As shown in Supporting Information Fig.~S8, such a modulation is evidenced by the replication of $\Gamma$-point features at the M-point. The Co-termination is overall the most prevalent and dominates high-resolution ARPES measurements with larger light spot sizes, as seen in the Supporting Information Figs.~S6-S12. In any case, our study reveals a bulk phenomenology, thus terminations do not affect the results we present. For simplicity, we accordingly restrict our analysis to data obtained from this termination and we move to the characterization of the magnetic phase.

\begin{figure}[htb]
\includegraphics[width=0.95\columnwidth]{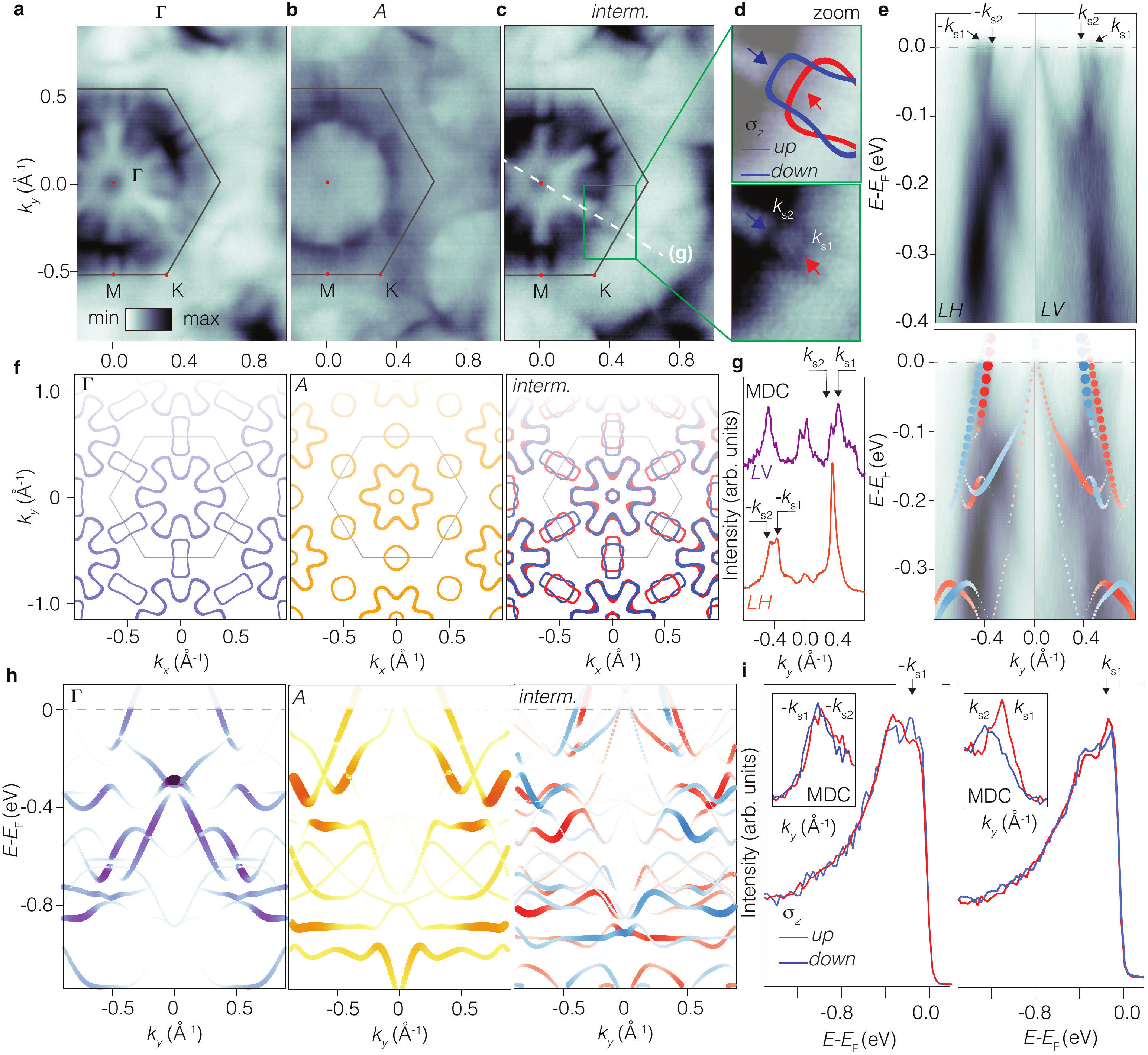}
  \caption{\textbf{Fermiology, DFT, and spin splitting.} Fermi surfaces at bulk $\Gamma$ (\SI{51}{\eV}, \textbf{a.}), A (\SI{75}{\eV}, \textbf{b.}), and an intermediate point near $k_z=\pi/2c$ (\SI{55}{\eV}, \textbf{c.}). \textbf{d.} Zoom-in on the splitting, with and without DFT calculations. The arrows indicate sections were the bands are prominently separated. \textbf{e.} \textit{Top}: (\textit{E}, \textit{k})-spectra at \SI{44}{\eV} (nominal $k_z$ near A; see Supporting Information, Section VIII), where matrix elements favor the split branches (see Supporting Information for additional energies), using \textit{p}- (LH, left) and \textit{s}-polarized (LV, right) light. \textit{Bottom}: same (\textit{E}, \textit{k})-spectra with overlaid DFT calculations. \textbf{f.} DFT spin-polarized 2D Fermi surfaces of CoNb$_4$Se$_8$ at $k_z=0$ , $k_z=\pi/c$, and $k_z=\pi/2c$.  \textbf{g.} MDCs at the Fermi level from \textbf{e.}, highlighting multiple split bands. \textbf{h.} DFT spin-polarized bands of CoNb$_4$Se$_8$ unfolded in the large Brillouin zone of ($1\times1$) NbSe$_2$ along $\mathrm{M}-\Gamma-\mathrm{M}$ for $k_z=0$, $k_z=\pi/c$, and $k_z=\pi/2c$, with spectral weight $P_{\vec{K}m}$ defined in Eq. \ref{eq:eq1}. \textbf{i.} spin-ARPES at opposite \textit{k}-points (\SI{55}{\eV} \textit{p}-polarized, as in \textbf{c.}), showing EDCs at $k_{s1}$ and MDCs at the Fermi level (insets).}
  \label{fig2}
\end{figure}

CoNb$_4$Se$_8$ crystallizes in the centrosymmetric space group P6$_3$/mmc, No. 194. The A-type collinear antiferromagnetic order along $c$ reduces the magnetic space group to P6$_3'$/m$'$m$'$c, and the corresponding non-relativistic spin space group is generated by combining the crystallographic operations of 6/mmm with the spin operations that relate the two opposite-spin Co sublattices, in agreement with the spin-group assignments of Refs.~\cite{Smejkal2022_1,Smejkal2022_2,Roig2024} for this class of compounds, and with the symmetry analyses recently reported on the same compound \cite{mandujano2025jacs,Dale_2024}. Operations that preserve the Co sublattices (the threefold rotations, inversion, the mirrors perpendicular to $\langle 100\rangle$ and the twofold axes along $\langle 100\rangle$) are combined with the identity in spin space, whereas those that exchange them (the $\{6_{001}|00\frac{1}{2}\}$ screw, the mirror $\{m_{001}|00\frac{1}{2}\}$, the $c$ glides perpendicular to $\langle 1\bar{1}0\rangle$ and the twofold axes along $\langle 1\bar{1}0\rangle$) are combined with a spin flip. In Litvin notation the spin space group is P$^{-1}$6$_3$/$^{-1}$m\,$^{1}$m\,$^{-1}$c\,$^{\infty m}$1, the same as for the NiAs-type altermagnets MnTe and CrSb \cite{Smejkal2022_1,Krempasky2024,Reimers_2024}; the full analysis is given in Supporting Information, Section VI and Table S3 \cite{Chen2024SSG,Xiao2024SSG,Jiang2024SSG}. From the general classification of space groups and their associated allowed altermagnetic spin-splittings \cite{Roig2024}, it follows that the lowest-order non-relativistic spin-split form factor compatible with the spin group of $2\times2$ reconstructed CoNb$_4$Se$_8$ is the $g$-wave harmonic
\begin{equation}
\Delta_s(\mathbf{k}) \propto k_z\, k_y\, (3k_x^2 - k_y^2),
\label{eq:gwave}
\end{equation}
whose nodal manifold consists of the two high-symmetry planes $k_z=0$ and $k_z=\pi/c$ together with the three vertical mirror planes containing the $\Gamma$--K high-symmetry lines. The splitting therefore vanishes in the $k_z=0$ and $k_z=\pi/c$ planes (as required by the antiferromagnetic Kramers degeneracy at those planes) and along the directions where the angular factor $k_y(3k_x^2-k_y^2)$ is zero, while it reaches its maximum at intermediate $k_z$ (such as $k_z=\pi/2c$) and at general $k$-points off the high-symmetry directions. The role of $k_z$ is thus pivotal: the breakdown of Kramers degeneracy manifests in the electronic structure away from bulk high-symmetry points. Photon-energy-dependent measurements are therefore required to select $k_z$, and fine photon-energy steps are used to tune the photoemission matrix elements and to select favorable $k_z$ windows. The large $c$-axis lattice constant of CoNb$_4$Se$_8$ (see Supporting Information Fig.~S1 and discussion) means each photon energy probes multiple $k_z$ values simultaneously, leading to partial spectral broadening \cite{Mitsuhashi2016}. Nevertheless, varying photon energy enables modulation of matrix elements and band intensities, which significantly enhances the detection of subtle spectral features. This capability is exemplified in the Fermi surface maps obtained at different photon energies, in Fig.~\ref{fig2}a--c (see also Supporting Information Figs.~S10--S12, where photon energy measurements with fine step are reported). Owing to the large $c$ axis, each photon energy integrates over a $k_z$ window ($\Delta k_z\approx\SI{0.2}{\angstrom^{-1}}$) comparable to the $\Gamma$--A distance ($\pi/c\approx\SI{0.25}{\angstrom^{-1}}$); the maps in Fig.~\ref{fig2}a--c are centred near bulk $\Gamma$, near A and at an intermediate $k_z$, respectively. The nodal structure is therefore identified through spin-resolved observables that are independent of $k_z$ (the null along $\Gamma$--K) or protected by symmetry against $k_z$ broadening (the reversal at opposite momenta; Supporting Information, Section VIII).

While all Fermi surfaces share common features, \textit{i.e.}, hexagonal flower-like central contours and elongated `spindles', the map in Fig.~\ref{fig2}c reveals a clear splitting in the electronic band structure, which complement and refine the picture emerging from prior ARPES work on the same compound \cite{Dale_2024,Sakhya2025}. In particular, the altermagnetic splitting is directly visible in our raw spin-integrated photoemission data, as resolved contours in the Fermi-surface maps at intermediate $k_z$ (Fig.~\ref{fig2}c, also highlighted by arrows in Fig.~\ref{fig2}d), and as two distinct peaks in the corresponding momentum distribution curves (MDCs) at the Fermi level (Fig.~\ref{fig2}g). The combination of micro-ARPES with a $\sim\SI{4}{\micro\metre}$ beam spot, fine photon-energy mapping in the range from \SIrange{25}{75}{\eV}, and systematic light-polarisation control further allows us to address individual single-termination regions, select favorable $k_z$ windows in which the \textit{g}-wave splitting is symmetry-allowed, and tune the photoemission matrix elements so that the otherwise nearly degenerate branches become unambiguously visible.

We note that in the Fermi surfaces, some of the spectral features appear as filled patches of intensity rather than contours as expected from density functional theory (DFT) calculations. This is attributable to photoemission matrix elements, which strongly vary with both geometry and photon energy. For the same reason, matrix element effects cause the splitting to be more or less evident at nominally the same $k_z=\pi/2c$, but different photon energies.

The experimental Fermi surfaces align with DFT (Fig.~\ref{fig2}f and Fig.~S13) in capturing the fermiology across the $k_z$ planes. The spin-splitting of the Kramers-paired bands at multiple momenta show remarkable agreement. This agreement is further highlighted in Fig.~\ref{fig2}d, where the spin-up (red) and spin-down (blue) bands from the DFT calculations coincide exactly with the experimental data. There is a similarly strong match between the DFT calculated (\textit{E}, \textit{k}) distributions across different $k_z$ values and acquired micro-ARPES data, as shown in Fig.~\ref{fig2}h, where blue and orange curves correspond to $\Gamma$ and A points, respectively (see also Supporting Information Fig.~S14 for DFT-calculated (\textit{E}, \textit{k}) spectra in the $\Gamma$--K direction). In order to display a simpler representation of the electronic dispersion for comparison with micro-ARPES measurements, we have mapped the spin-polarized electronic eigenvalues and eigenvectors of the bulk altermagnetic $2\times2$ CoNb$_4$Se$_8$ supercell (SC) into the bigger Brillouin zone of the $1\times1$ NbSe$_2$ primitive cell (PC) via an unfolding scheme~\cite{PhysRevB.85.085201}. The thickness of the bands in the unfolded electronic dispersions is proportional to the spectral weight:

\begin{equation}\label{eq:eq1}
P_{\vec{K}m}(\vec{k}_i)=\sum_{n}\biggl|\braket{\vec{K}m}{\vec{k}_in}\biggr|^2,
\end{equation}

\noindent which is the probability of finding a set of PC states, $\ket{\vec{k}_in}$, contributing to the SC state, $\ket{\vec{K}m}$~\cite{PhysRevB.85.085201}, where $m$ and $n$ are band indices and $i$ runs over the number of $\vec{K}$ vectors.

Notably, while eigenvalues for $k_z=0$ and $k_z=\pi/c$ are spin-degenerate (as expected for an antiferromagnetic system), DFT results for $k_z=\pi/2c$ (rightmost panel in Fig.~\ref{fig2}h, where the red/blue colors are used to highlight spin-up/down bands) provide a precise benchmark for finding where Kramers degeneracies are lifted, identifying CoNb$_4$Se$_8$ as a \textit{g}-wave altermagnet and enabling a one-to-one correspondence with the experimental data. Near the Fermi level, the theoretical altermagnetic splitting is in excellent agreement with the experimental spectra (lower panel in Fig.~\ref{fig2}e), and the bands forming the altermagnetic pairs are labeled as $\pm k_{s1,s2}$. Including spin--orbit coupling (SOC) opens gaps of up to \SI{0.1}{\eV} and \SI{0.2}{\eV} at band crossings at $\Gamma$ and A, respectively, where the altermagnetic splitting vanishes by symmetry. At the momenta of the altermagnetic pairs $\pm k_{s1,s2}$, SOC shifts the bands by less than \SI{20}{\milli\eV} and leaves $\langle S_z\rangle\approx\pm0.5\hbar$ (essentially fully polarized), whereas the nonrelativistic splitting is $\approx\SI{150}{\milli\eV}$ (Supporting Information, Section VI and Table S4). Moreover, any time-reversal-even $\sigma_z$, including the hidden Ising polarization of the NbSe$_2$ layers \cite{Zhang2014hidden,Riley2014}, is forbidden along $\Gamma$--M by the vertical mirror containing this direction, and cannot account for the spin-ARPES pattern discussed below. The non-zero spin character of these bands is further validated by spin-ARPES measurements (Fig.~\ref{fig2}i), collected for the $\sigma_z$ spin-component at multiple \textit{k} values within the plane centered at $k_z=\pi/2c$. Both energy distribution curves (EDCs) at $\pm k_{s1}$ and MDCs at the Fermi level across numerous $k$ points reveal a small, yet clear, spin-reversal pattern. The corresponding spin-polarization and its $1\sigma$ statistical error bands are shown in Supporting Information Fig.~S23, to directly assess the significance of the sign reversal. The antisymmetric part of the polarization, $[P(k)-P(-k)]/2\approx6\pm1.5\%$ in magnitude at $E-E_F\approx\SI{-0.05}{\eV}$, corresponds to about four standard deviations and is insensitive by construction to any $k$-independent instrumental offset; the data treatment is detailed in Supporting Information, Section VII. This behavior is consistent with theoretical predictions and is observed at four different $k$ points in the MDCs (insets of Fig.~\ref{fig2}i), further supporting the validity of our findings. Importantly, a non-zero spin signal and sign swap at time-reversal momenta have been found systematically for various cleaves. Moreover, moving the beam spot by $\approx\SI{1}{\milli\metre}$ on the same cleaved surface reverses the spin contrast at the same momentum, consistent with a region of opposite time-reversal domain (Supporting Information, Fig.~S20). An additional sanity check is provided by spin-ARPES along the $\Gamma$--K direction, which corresponds to a symmetry-protected nodal plane of the \textit{g}-wave form factor of Eq.~\eqref{eq:gwave}: the measured spin signal along this direction is compatible with zero within the $1\sigma$ uncertainty (Supporting Information, Figs.~S19 and S23). In the bulk, $\sigma_z$ along $\Gamma$--K vanishes by symmetry even when SOC is included, because the $c$ glide containing $\Gamma$--K is a symmetry of P6$_3'$/m$'$m$'$c without time reversal; at the surface this glide is lost, but the hidden Ising polarization of the topmost NbSe$_2$ layer that would then be allowed is expected to be strongly reduced by the covalent bonding of adjacent layers through the intercalated Co (Supporting Information, Section VII). This discussion, along with considerations about the domain size and extra sets of data, is expanded in the dedicated section VII of the Supporting Information and Figs.~S19 to S23. The combination of micro-ARPES and spin-ARPES here presented thus robustly confirms the altermagnetic character of CoNb$_4$Se$_8$, in excellent agreement with DFT calculations showing the \textit{g}-wave altermagnetic splitting and its $k_z$ dependence.

\begin{figure}
\includegraphics[width=1.0\columnwidth]{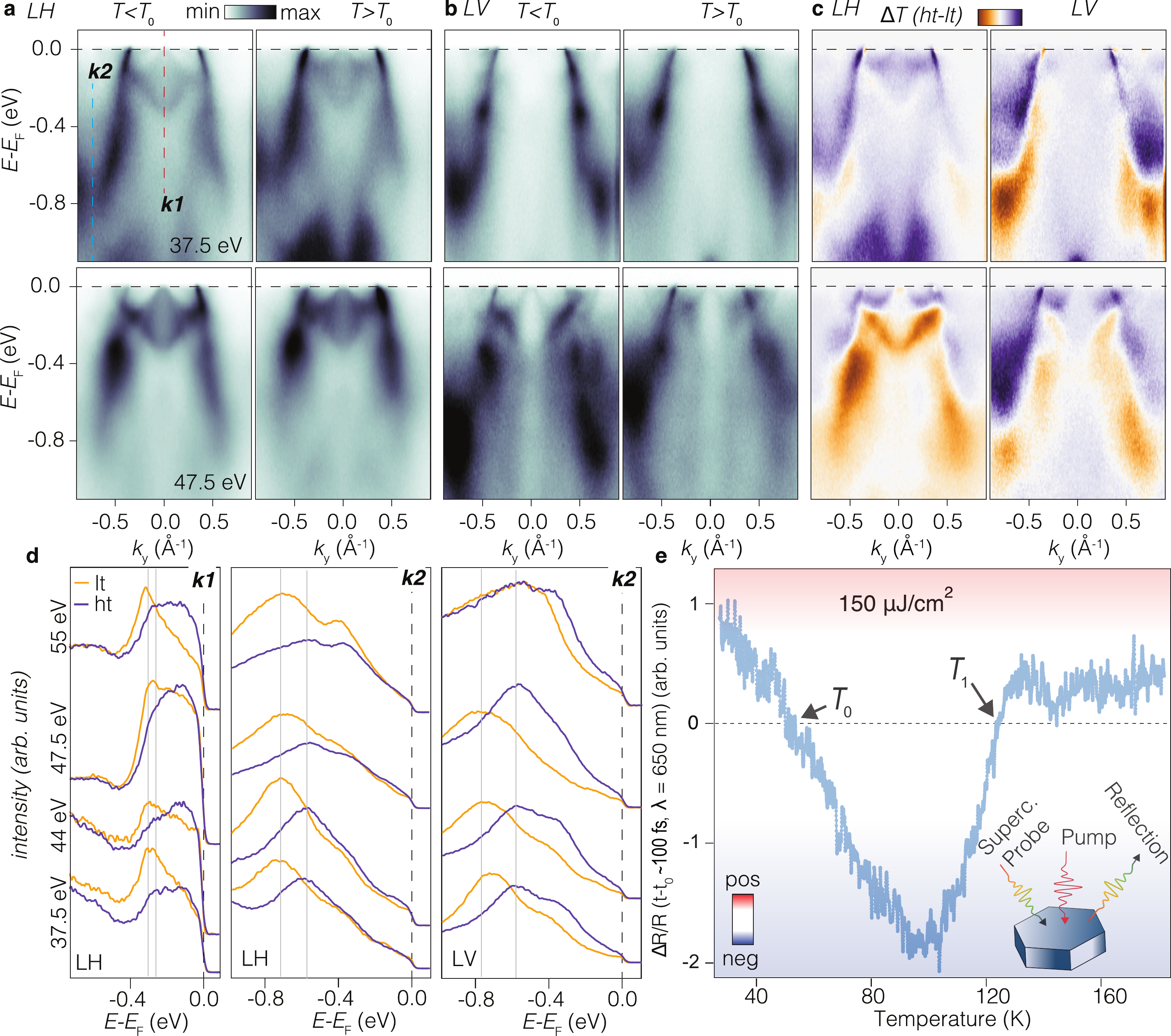}
   \caption{\textbf{Low-temperature anomaly.} ARPES spectra (black indicates high intensity in the ARPES scale) collected at two distinct photon energies:  \SI{37.5}{\eV} (top row) and \SI{47.5}{\eV} (bottom row), and for temperatures above and below $T_0$. These are shown for both \textbf{a.} \textit{p}-polarization (LH) and \textbf{b.} \textit{s}-polarization (LV). \textbf{c.} Difference between spectra collected above and below $T_0$; orange (purple) corresponds to a negative (positive) difference. The strong electronic renormalization manifests as a decrease in the overall bandwidth, while $k_F$ remains constant. \textbf{d.} EDCs collected at different photon energies (indicated along the vertical axis) for spectra below $T_0$ (orange) and above it (purple). These are shown for two \textit{k} values (\textit{k1} and \textit{k2}) as indicated in \textbf{a.} \textbf{e.} $\Delta R/R$ signal as measured by ultrafast reflectivity at $t-t_0\sim\SI{100}{\femto\s}$ for \SI{150}{\micro\J\per\square\cm} fluence. The two zero-crossings of the $\Delta R/R$ signal at $T_0\approx\SI{50}{\K}$ and $T_1\approx\SI{130}{\K}$ are highlighted. The temperature axis refers to the sensor; on the sample, $T_0$ corresponds to $\sim\SI{60}{\K}$ and $T_1$ corresponds to $\sim\SI{170}{\K}$.}
    \label{fig3}
\end{figure}

Interestingly, we observe a small kink at $T\approx\SI{50}{\K}$ in the magnetization as measured by SQUID (Supporting Information Fig.~S3), which we refer to as the low-temperature anomaly at $T_0$. We note that this temperature does not coincide with the $\sim\SI{28}{}$--$\SI{36}{\K}$ transition recently associated with the non-centrosymmetric $\sqrt{3}\times\sqrt{3}\,R30^\circ$ ordering of Co-excess CoNb$_{3}$Se$_6$ \cite{Regmi2024,mandujano2025jacs,mandujano2025prm,graham2026}, which is not detected in our diffraction probes (Supporting Information Fig.~S1). Moreover, due to competing magnetic interactions, even small changes in the Co concentration would stabilize a magnetic transition temperature lower than \SI{170}{ \K} \cite{mandujano2025jacs}, which is inconsistent with our data. Therefore, the presence of this signal does not change the robustness of the conclusions presented above, however it warrants further investigation. In order to get more insight into this phenomenon, we measured micro-ARPES spectra below and above $T_0$ at two photon energies and at linear vertical (LV) and linear horizontal (LH) light polarizations. We report a temperature-driven behavior (Fig.~\ref{fig3}a), with a significant band renormalization (Fig.~\ref{fig3}a--d and Supporting Information Fig.~S7). We observe the renormalization across multiple photon energies and independently of light polarization; in particular, Fig.~\ref{fig3}c displays the difference in spectral intensity between data collected above and below $T_0$ for the different photon energies and polarizations, so that modifications to the bandwidth appear more evidently. This is also shown more explicitly in Fig.~\ref{fig3}d with EDCs extracted at specific \textit{k} points (indicated in Fig.~\ref{fig3}a) across various photon energies for both temperatures. Crucially, micro-ARPES provides precise spatial control, allowing us to maintain the same position when varying the temperature.

Additionally, we performed time-resolved reflectivity measurements on the same batch of samples. Fig.~\ref{fig3}e shows the evolution of the reflectivity variation ($\Delta R/R$) measured shortly after the absorption of the pump pulse ($t-t_0 \sim \SI{100}{\femto\s}$). At a fluence of $\approx\SI{150}{\micro\J\per\square\cm}$, two distinct sign changes in the $\Delta R/R$ signal are observed near \SI{130}{\K} and \SI{50}{\K}, as measured directly by the cryostat sensor, corresponding to $\sim\SI{170}{\K}$ and $\sim\SI{60}{\K}$ on the sample due to nonlinear sensor-to-sample offset. A sign change of $\Delta R/R$ is commonly observed when the electronic structure changes at the onset of an ordered phase, as in cuprate superconductors and charge-density-wave systems \cite{shimizu2025,wang2025}. The sign change near \SI{130}{\K} can be related to the paramagnetic-altermagnetic transition; the effect near \SI{50}{\K} coincides with the band renormalization shown in the micro-ARPES spectra in Fig.~\ref{fig3} and with the kink in the magnetization.

The microscopic origin of this anomaly remains open, and extrinsic effects cannot be excluded. We only note that the optical response is electronic in character: as detailed in the Supporting Information, Section IV, the steady-state pump-induced heating is below \SI{3}{\K}, and the reflectivity signal $\Delta R/R \; (\SI{650}{\nm})$ evolves on a sub-\SI{200}{\femto\s} timescale, faster than expected for phonon-mediated effects. This statement concerns the probe signal and not the origin of the anomaly: the sign change reflects a modification of the allowed optical transitions, consistent with the band renormalization observed by micro-ARPES. No anomaly is resolved in the heat capacity near $T_0$ (Supporting Information, Section I); given the dominant lattice contribution at these temperatures, this does not exclude a small electronic contribution, and high-resolution thermodynamic measurements will be required to establish the origin of the anomaly.

In conclusion, we have resolved the altermagnetic band splitting of the layered \textit{g}-wave altermagnet CoNb$_4$Se$_8$, stabilized via Co intercalation in 2H-NbSe$_2$. Using micro-ARPES, spin-ARPES, and temperature-dependent reflectivity combined with DFT, we provide direct evidence of Kramers degeneracy lifting and reveal the spin-polarized band splitting characteristic of altermagnetic phenomena; moreover, we report a low-temperature anomaly accompanied by a renormalization of the bandwidth with temperature, whose microscopic origin requires further investigation. These findings establish CoNb$_4$Se$_8$ as a well-characterized model system for layered altermagnets, on which future studies of exfoliated flakes, heterostructures and transport can build.

\noindent{\bf Acknowledgements}\\
F.M. greatly acknowledges the NFFA-DI funded by the European Union – NextGenerationEU, M4C2, within the PNRR project NFFA-DI, Grants No. CUP B53C22004310006 and No. IR0000015, and MUR funding within the FIS2 (n. 1236, 1-8-2023) Project no. FIS-2023-03652 (CUP D53C25001950001). A.D.V. acknowledges funding from the Deutsche Forschungsgemeinschaft (DFG) within Transregio TRR 227 Ultrafast Spin Dynamics project B07, the Max Planck Society, and BERLIN QUANTUM, an initiative endowed by the Innovation Promotion Fund of the city of Berlin. D.R. acknowledges support from the HPC resources of IDRIS, CINES, and TGCC under Allocation No. 2024-A0160914101 made by GENCI. M.B.P. and J.A.M gratefully acknowledge support from the Danish Agency for Science and Higher Education (DanScatt Grant No. 7129-00018B). M.C. acknowledges the European Union (ERC, DELIGHT, 101052708). I.Z. gratefully acknowledges the support from the US Department of Energy grant number DE-SC0025005. We thank A. Jones for his support during  measurements at the SGM4 beamline. This work was partially performed in the framework of the Nanoscience Foundry and Fine Analysis (NFFA-MUR Italy Progetti Internazionali) project. F. Motti acknowledges the support of the EC project SINFONIA (Grant No. H2020-FET-OPEN-964396). This work was supported by the National Research Foundation of Korea (NRF) funded by the Ministry of Education, Science and Technology (Grants No. NRF-2019M2C8A1057099 and No. NRF-2022R1I1A1A01063507). J.S.K. was supported by the Institute for Basic Science (IBS) through the Center for Artificial Low-Dimensional Electronic Systems (IBS-R014-D1). Y.H.H. was supported by the National Research Foundation of Korea (NRF) grant funded by the Korean government (MSIT) (RS-2025-24742993). The authors gratefully thank Prof. I. Mazin and Prof. N.J. Ghimire for useful insights and discussion.

\noindent{\bf Author contributions}\\
F.M., M.C. coordinated the project. F.M., A.D.V., M.C., D.R., J.A.M, and I.Z. wrote the manuscript with input from all authors.  F.M., A.D.V., C.B., M.D.W., F.B., O.J.C, J.A.M., I.V., J.F., M.Z., M. B. P., and J.B.J. performed the ARPES measurements. F.Mo. and G.V. performed high-resolution XPS measurements. F.C. and M.T. performed the optical spectroscopy, M.C. and D.R. performed the DFT calculations, I.Z., C.C., M. X., S.C., A.L. performed the STM measurements, and Y.H. synthesized the crystals, performed the resistivity, and initial characterization. W.S.S. and J.S.K. performed and analyzed the heat-capacity measurements. All authors contributed to the discussion of the work. 


\noindent{\bf Methods}\\

\noindent{\bf Sample Growth}\\
Single crystals of CoNb$_4$Se$_8$ were synthesized using the chemical vapor transport (CVT) method. High-purity cobalt (Co, 99.99$\%$), niobium (Nb, 99.999$\%$), and selenium (Se, 99.9999$\%$) powders served as starting materials. To prevent contamination and residual oxygen during synthesis, the quartz ampoule was chemically cleaned and subjected to vacuum heat treatment prior to loading. The raw materials were then sealed in a quartz ampoule (approximately 10\,mm in diameter and 150\,mm in length) along with iodine (5\,mg/cm$^3$ relative to ampoule volume) as the transport agent. The sealed ampoule was evacuated to a high vacuum and placed in a two-zone horizontal furnace, with the source region maintained at a higher temperature than the growth region. Optimal temperature gradients and iodine concentration were critical for achieving high-quality crystal growth. For CoNb$_4$Se$_8$, the source region was held at 960–980\,$^\circ$C, while the growth region temperature was initially increased incrementally from 880\,$^\circ$C to 900\,$^\circ$C over 100 hours. Subsequently, both zones were held at constant temperatures for an additional 300 hours to facilitate the growth of large single crystals. Finally, the temperature was reduced gradually over 100 hours, with the source region cooled to 200\,$^\circ$C and the growth region to 100\,$^\circ$C, before allowing the ampoule to cool naturally to room temperature. The resulting single crystals measured approximately 5 $\times$ 5 $\times$ 0.1\,mm$^3$. Residual iodine was removed from the crystals by rinsing with a methanol solution. The mole fraction `\textit{x}' in Co$_x$NbSe$_2$ was preliminarily determined using energy-dispersive x-ray spectroscopy (EDS) in a field-emission scanning electron microscope (FE-SEM, JEOL 7500). High-precision quantification was subsequently performed using x-ray photoelectron spectroscopy (XPS) at the APE-HE beamline of the Elettra synchrotron radiation facility in Trieste, Italy (details in Supporting Information). The crystal structure and quality were confirmed using high-resolution X-ray diffraction (XRD, D8 Advance, Bruker, Germany). Additional sample characterization is shown in sections I and II of the Supporting Information.

\noindent{\bf Synchrotron radiation measurements}\\
Micro-ARPES measurements were performed on in situ cleaved samples. The beam was focused using the capillary mirror optic at the I05 beamline at Diamond Light Source, with a final beam spot of $\sim\SI{4}{\micro\metre}$ FWHM. The base sample temperature was \SI{25}{\K}. The standard ARPES data were acquired at the CASSIOPEE beamline at Synchrotron SOLEIL (Paris, France) with momentum and energy resolutions better than \SI{0.018}{\angstrom^{-1}} and \SI{10}{\milli\eV}, respectively. The sample temperature was fixed at circa \SI{20}{\K}, \textit{i.e.} well below the magnetic transition temperature, for all measurements. The Fermi surfaces were obtained by rotating the sample around the analyzer focus, with the slit positioned orthogonal to the rotation axis. Data were acquired using photon energies in the range from \SIrange{25}{75}{\eV}, which allows us to tune the photoemission matrix elements and to locate the high-symmetry points along $k_z$ (Supporting Information, Section VIII). Spin-ARPES was performed using V-LEED technology at the APE-LE laboratory at the Elettra Synchrotron (Trieste, Italy) and with LH light. For these measurements, the samples were aligned with the slit along the $\Gamma-\mathrm{M}$ direction and kept under normal emission conditions. The various momenta were reached via use of deflectors. The spin-ARPES was performed with the sample held at a temperature of \SI{15}{\K}. The treatment of the spin-resolved data is described in Supporting Information, Section VII. In all ARPES and spin-ARPES measurements, the samples were fixed to the sample holder using silver epoxy (HD20E, Epotek). A small ceramic post was secured to the top of the sample using the same epoxy. Once the samples were introduced into the measurement chamber, they were cleaved under ultrahigh vacuum conditions (better than $\SI{1e-10}{\milli\barpressure}$) and after being cooled at the base temperature of \SI{15}{\K}.

\noindent{\bf Computational details}\\
Collinear spin-polarized DFT calculations (without Hubbard correction, \textit{i.e.}, DFT+\textit{U}) have been performed using the plane-wave pseudopotential method as implemented in the Quantum ESPRESSO package v.7.2~\cite{QE-2009,QE-2017}. We employed a norm-conserving pseudopotential for Se atoms and ultrasoft pseudopotentials for Nb and Co atoms, the latter taken from the Garrity--Bennett--Rabe--Vanderbilt library \cite{GARRITY2014}, with an energy cut-off of $45$ Ry and $450$ Ry for the wave-function and electron density respectively. Electronic band structure and energetics have been obtained through Perdew-Burke-Ernzerhof (PBE)~\cite{PhysRevLett.77.3865} exchange-correlation functional. The Brillouin zone has been sampled with a k-vector $\Gamma$-centered Monkhorst–Pack mesh of $9\times9\times16$ points and a first-order Methfessel-Paxton~\cite{PhysRevB.40.3616} electronic smearing of $5$ mRy, assuring convergence of the total energy per atom in the supercell within \SI{1}{\milli\eV}/atom. We investigated the energetics of Co-intercalated $2\times2$ NbSe$_2$ (CoNb$_4$Se$_8$) in the paramagnetic (PM), ferromagnetic (FM) and altermagnetic (ALM) configuration and as a function of the Co position by relaxing both lattice parameters and internal coordinates until residual forces were below 10$^{-3}$ Ry/Bohr (see Supporting Information, Section~ VI). Bulk altermagnetic $2\times2$ CoNb$_4$Se$_8$ supercell eigenvalues and eigenvectors have then been mapped into the $1\times1$ NbSe$_2$ primitive-cell Brillouin zone using the scheme of Popescu and Zunger~\cite{PhysRevB.85.085201} and implemented in the code unfold.x~\cite{10.1063/5.0047266}.

\noindent{\bf Extended data} is available for this paper at https://xxxx

\noindent{\bf Supporting Information} The online version contains Supporting Information available at https://xxx

\noindent{\bf Data availability} The data that support the findings of this study are available from the corresponding authors upon reasonable request.

\noindent{\bf Code availability} The code that support the findings of the study is available from the corresponding authors on reasonable request.

\noindent{\bf Correspondence and requests for materials} should be addressed to \href{mailto:devita@fhi-berlin.mpg.de}{A. De Vita}, \href{mailto:chiara.bigi@synchrotron-soleil.fr}{C. Bigi}, \href{mailto:ilija.zeljkovic@bc.edu}{I. Zeljkovic}, \href{mailto:younghh@uc.ac.kr}{Y. Hwang}, \href{mailto:m.calandrabuonaura}{M. Calandra}, \href{mailto:miwa@phys.au.dk}{J. Miwa}, \href{mailto:federico.mazzola@spin.cnr.it}{F. Mazzola}.

\noindent{\bf Ethics declarations} The authors declare no competing interests.

\bibliographystyle{MSP}

\bibliography{alter.bib}

\end{document}


\title{Supporting Information: Robust spin splitting and fermiology in the layered altermagnet CoNb$_4$Se$_8$}
	
	\author{Alessandro De Vita***}\email{devita@fhi-berlin.mpg.de}
	\affiliation{Institut für Physik und Astronomie, Technische Universität Berlin, Straße des 17 Juni 135, 10623 Berlin, Germany}
	\affiliation{Fritz Haber Institut der Max Planck Gesellshaft, Faradayweg 4--6, 14195 Berlin, Germany}
	
	\author{Chiara Bigi***}\email{chiara.bigi@synchrotron-soleil.fr}
	\affiliation{Synchrotron SOLEIL, F-91190 Saint-Aubin, France}
	
	\author{Davide Romanin***}
	\affiliation{Université Paris-Saclay, CNRS, Centre de Nanosciences et de Nanotechnologies, 91120, Palaiseau, Paris, France}
	
	\author{Matthew D. Watson}
	\affiliation{Diamond Light Source, Harwell Campus, Didcot, OX11 0DE, United Kingdom}
	
	\author{Vincent Polewczyk}
	\affiliation{Universit\'e Paris-Saclay, UVSQ, CNRS, GEMaC, 78000, Versailles, France}
	
	\author{Marta Zonno}
	\affiliation{Synchrotron SOLEIL, F-91190 Saint-Aubin, France}
	
	\author{Fran\c cois Bertran}
	\affiliation{Synchrotron SOLEIL, F-91190 Saint-Aubin, France}
	
	\author{Pietro M. Forcella}
	\affiliation{Université Paris-Saclay, CNRS, Centre de Nanosciences et de Nanotechnologies, 91120, Palaiseau, Paris, France}
	
	\author{My Bang Petersen}
	\affiliation{Department of Physics and Astronomy, Interdisciplinary Nanoscience Center, Aarhus University, 8000 Aarhus C, Denmark}
	
	\author{Federico Motti}
	\affiliation{CNR-Istituto Officina dei Materiali (IOM), Unità di Trieste, Strada Statale 14, km 163.5, 34149 Basovizza (TS), Italy\looseness=-1}%
	
	\author{Giovanni Vinai}
	\affiliation{CNR-Istituto Officina dei Materiali (IOM), Unità di Trieste, Strada Statale 14, km 163.5, 34149 Basovizza (TS), Italy\looseness=-1}%
	
	\author{Manuel Tuniz}
	\affiliation{Dipartimento di Fisica, Università degli Studi di Trieste, 34127, Trieste, Italy}
	
	\author{Federico Cilento}
	\affiliation{Elettra - Sincrotrone Trieste S.C.p.A., Strada Statale 14, km 163.5, Trieste, Italy}
	
	\author{Mario Cuoco}
	\affiliation{CNR-SPIN, c/o Universit\'a di Salerno, IT-84084 Fisciano (SA), Italy}
	
	\author{Brian M. Andersen}
	\affiliation{Niels Bohr Institute, University of Copenhagen, 2100 Copenhagen, Denmark}
	
	\author{Andreas Kreisel}
	\affiliation{Niels Bohr Institute, University of Copenhagen, 2100 Copenhagen, Denmark}
	
	\author{Luciano Jacopo D'Onofrio}
	\affiliation{CNR-SPIN, c/o Universit\'a di Salerno, IT-84084 Fisciano (SA), Italy}
	
	\author{Oliver J. Clark}
	\affiliation{School of Physics and Astronomy, Monash University, Clayton, Victoria 3800, Australia}
	
	\author{Mark T. Edmonds}
	\affiliation{School of Physics and Astronomy, Monash University, Clayton, Victoria 3800, Australia}
	
	\author{Christopher Candelora}
	\affiliation{Department of Physics, Boston College, Chestnut Hill, MA 02467, USA}
	
	\author{Muxian Xu}
	\affiliation{Department of Physics, Boston College, Chestnut Hill, MA 02467, USA}
	
	\author{Siyu Cheng}
	\affiliation{Department of Physics, Boston College, Chestnut Hill, MA 02467, USA}
	
	\author{Alexander LaFleur}
	\affiliation{Department of Physics, Boston College, Chestnut Hill, MA 02467, USA}
	
	\author{Tommaso Antonelli}
	\affiliation{ETH Zürich, HPF E 19 Otto-Stern-Weg 1, 8093 Zürich, Switzerland}
	
	\author{Giorgio Sangiovanni}
	\affiliation{Institute for Theoretical Physics and Astrophysics, University of W\"urzburg, D-97074 W\"urzburg, Germany}
	
	\author{Lorenzo Del Re}
	\affiliation{Institute for Theoretical Physics and Astrophysics, University of W\"urzburg, D-97074 W\"urzburg, Germany}
	
	\author{Ivana Vobornik}
	\affiliation{CNR-Istituto Officina dei Materiali (IOM), Unità di Trieste, Strada Statale 14, km 163.5, 34149 Basovizza (TS), Italy\looseness=-1}%
	
	\author{Jun Fujii}
	\affiliation{CNR-Istituto Officina dei Materiali (IOM), Unità di Trieste, Strada Statale 14, km 163.5, 34149 Basovizza (TS), Italy\looseness=-1}%
	
	\author{Fabio Miletto Granozio}
	\affiliation{CNR-SPIN, c/o Complesso di Monte S. Angelo, IT-80126 Napoli, Italy}
	
	\author{Alessia Sambri}
	\affiliation{CNR-SPIN, c/o Complesso di Monte S. Angelo, IT-80126 Napoli, Italy}
	
	\author{Emiliano Di Gennaro}
	\affiliation{Physics Department, University of Napoli “Federico II”, Via Cinthia, 21, Napoli 80126, Italy}
	
	\author{Jeppe B. Jacobsen} 
	\affiliation{Nanoscience Center, Niels Bohr Institute, University of Copenhagen, 2100 Copenhagen, Denmark}
	
	\author{Henrik Jacobsen} 
	\affiliation{European Spallation Source ERIC - Data Management and Software Center, 2800 Kgs.~Lyngby, Denmark}
	
	\author{Iulia Cojocariu}
	\affiliation{Elettra - Sincrotrone Trieste S.C.p.A., Strada Statale 14, km 163.5, Trieste, Italy}
	\affiliation{Dipartimento di Fisica, Università degli Studi di Trieste, 34127, Trieste, Italy}
	
	\author{Marcin Szpytma}
	\affiliation{Elettra - Sincrotrone Trieste S.C.p.A., Strada Statale 14, km 163.5, Trieste, Italy}
	\affiliation{Faculty of Physics and Applied Computer Science, AGH University of Krakow, Krakow, Poland}
	
	\author{Andrea Locatelli}
	\affiliation{Elettra - Sincrotrone Trieste S.C.p.A., Strada Statale 14, km 163.5, Trieste, Italy}
	
	\author{Tevfik Mentes}
	\affiliation{Elettra - Sincrotrone Trieste S.C.p.A., Strada Statale 14, km 163.5, Trieste, Italy}
	
	\author{Won Seop Song}
	\affiliation{Department of Physics, Pohang University of Science and Technology, 77 Cheongam-ro, Nam-gu, Pohang-si, Gyeongsangbuk-do, Korea 37673}
	\affiliation{Center for Artificial Low-Dimensional Electronic Systems, Institute for Basic Science, Campus 79, Jigok-ro 127beon-gil, Namgu, Pohang, Korea, (37673)}
	
	\author{Jun Sung Kim}
	\affiliation{Department of Physics, Pohang University of Science and Technology, 77 Cheongam-ro, Nam-gu, Pohang-si, Gyeongsangbuk-do, Korea 37673}
	\affiliation{Center for Artificial Low-Dimensional Electronic Systems, Institute for Basic Science, Campus 79, Jigok-ro 127beon-gil, Namgu, Pohang, Korea, (37673)}
	
	\author{Matthieu Jamet}
	\affiliation{Univ. Grenoble Alpes, CEA, CNRS, IRIG-SPINTEC, 38000 Grenoble, France}
	
	\author{Jean-François Jacquot}
	\affiliation{Univ. Grenoble Alpes, CEA, CNRS, IRIG-SPINTEC, 38000 Grenoble, France}
	
	\author{Pasquale Orgiani}
	\affiliation{CNR-Istituto Officina dei Materiali (IOM), Unità di Trieste, Strada Statale 14, km 163.5, 34149 Basovizza (TS), Italy\looseness=-1}%
	
	\author{Ralph Ernstorfer}
	\affiliation{Institut für Physik und Astronomie, Technische Universität Berlin, Straße des 17 Juni 135, 10623 Berlin, Germany}
	\affiliation{Fritz Haber Institut der Max Planck Gesellshaft, Faradayweg 4--6, 14195 Berlin, Germany}
	
	\author{Ilija Zeljkovic}\email{ilija.zeljkovic@bc.edu}
	\affiliation{Department of Physics, Boston College, Chestnut Hill, MA 02467, USA}
	
	\author{Younghun Hwang}\email{younghh@uc.ac.kr}
	\affiliation{Electricity and Electronics and Semiconductor Applications, Ulsan College, Ulsan 44610, Republic of Korea}
	
	\author{Matteo Calandra}\email{m.calandrabuonaura@unitn.it}
	\affiliation{Department of Physics, University of Trento, Via Sommarive 14, Povo 38123, Italy}
	
	\author{Jill A. Miwa}\email{miwa@phys.au.dk}
	\affiliation{Department of Physics and Astronomy, Interdisciplinary Nanoscience Center, Aarhus University, 8000 Aarhus C, Denmark}
	
	\author{Federico Mazzola}\email{federico.mazzola@unipd.it}
	\affiliation{CNR-SPIN, c/o Complesso di Monte S. Angelo, IT-80126 Napoli, Italy}

	\maketitle
	\noindent
	*** These authors contributed equally.
	
	High-purity CoNb$_4$Se$_8$ single crystals, synthesized as detailed in the Methods Section of the main text, underwent comprehensive characterization by X-ray diffraction (XRD), low-energy electron diffraction (LEED), resistivity measurements, X-ray photoelectron spectroscopy (XPS), pump-probe optical spectroscopy, and scanning tunnelling microscopy (STM). These complementary analyses established the structural and electronic behaviour of the samples, underpinning investigations using high-resolution angle-resolved photoelectron spectroscopy (ARPES), ARPES with few-micron spatial resolution resolution (micro-ARPES) and spin-resolved ARPES (spin-ARPES). In this section, we present an in-depth detailed discussion of the characterization outcomes, alongside additional ARPES-type data that support the findings of the main text.
	
	\subsubsection*{\textbf{I. Structural characterization of CoNb$_4$Se$_8$}}
	
	In this section, we present structural characterization and basic transport measurements of CoNb$_4$Se$_8$. XRD measurements confirm a clean and well-ordered crystalline structure consistent with the NbSe$_2$ lattice. Using data acquired with a D8 Advance Bruker diffractometer, we applied Bragg's law to analyze the positions of the intense diffraction peaks -- 002, 004, 006, and 008 -- shown in Fig. \ref{figS1}a. From this analysis, we extracted an out-of-plane (c-axis) lattice parameter of \SI{12.59}{\angstrom}, in good agreement with the step-edge found by microscopy (See Fig.~\rev{S3}).
	
	\renewcommand{\thefigure}{S1}
	\begin{figure}[htb]
		\includegraphics[width=1.0\textwidth]{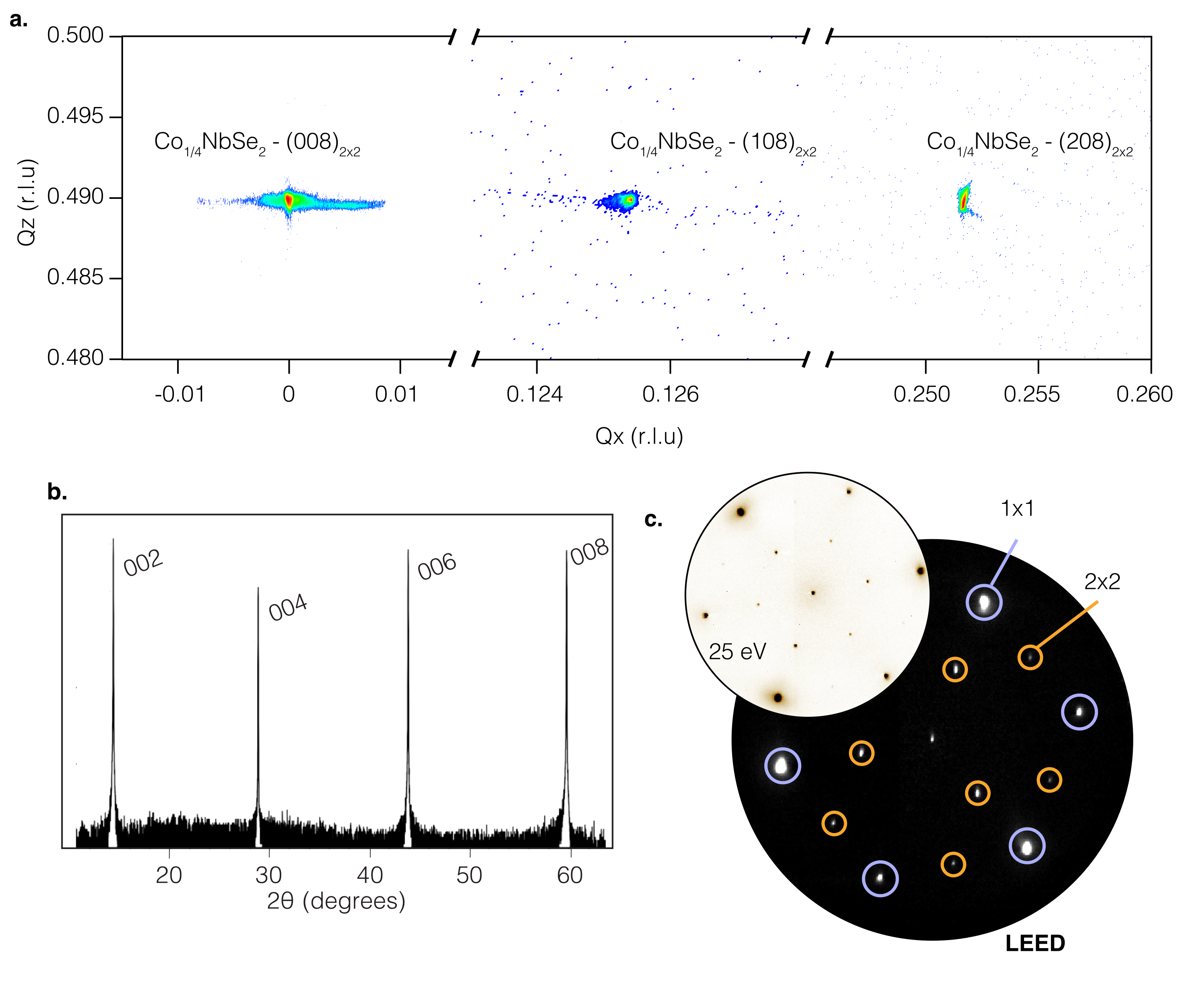}
		\caption{\textbf{Crystal structure properties of CoNb$_4$Se$_8$}. \textbf{a.} Asymmetric XRD measurements of the CoNb$_4$Se$_8$ revealing a clear 2$\times$2 order. The measurements have been performed at room temperature, indicating that the reconstruction is a peculiarity of the bulk system. \textbf{b} XRD measurements reveal a \textit{c}-axis lattice parameter of circa \SI{12.5}{\angstrom}. \textbf{c.} LEED data acquired at \SI{25}{\eV} in the altermagnetic state also showing the reconstruction: the orange circles indicate the 2$\times$2 features, while the blue circles mark the primitive 1$\times$1 ones.}
		\label{figS1}
	\end{figure}
	
	In CoNb$_4$Se$_8$, the Co atoms double the unit cell, giving a 2$\times$2 modulation. Such atoms are positioned between the single NbSe$_2$ layers, with the c-axis lattice parameter determined by the XRD measurements. The magnetic moments of the Co atoms are represented by red and blue arrows in the schematic and give rise to an alternating up-down stacking along the c-axis, which yields an overall zero magnetic moment (See also Fig.~S2 for magnetic characteristics). The 2$\times$2 modulation is also observed by LEED measurements: sharp first-order spots are associated with NbSe$_2$ units, which are clearly arranged in a hexagonal pattern and denoted by blue circles, and a 2$\times$2 pattern on top, highlighted by orange circles. Both our bulk (XRD) and surface (LEED) diffraction probes show exclusively the $2\times2$ superstructure on top of the primitive $1\times1$ NbSe$_2$ pattern, with no detectable $\sqrt{3}\times\sqrt{3}\,R30^\circ$ reflections; the latter would be the diagnostic fingerprint of the CoNb$_{3}$Se$_6$-type sublattice \cite{mandujano2025jacs,mandujano2025prm,Parkin1983,VanLaar1971}, and their absence is consistent with the stoichiometric CoNb$_4$Se$_8$ phase being the dominant ordered structure in our samples.
	
	\renewcommand{\thefigure}{S2}
	\begin{figure}[htb]
		\includegraphics[width=.6\textwidth]{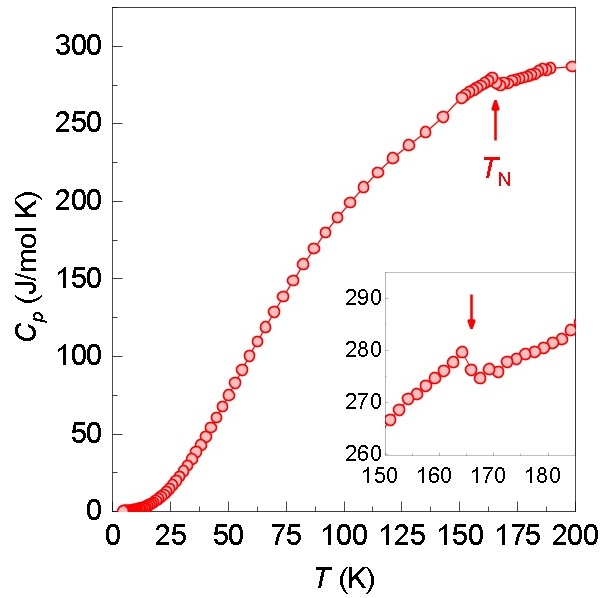}
		\caption{\textbf{Heat capacity}. Temperature-dependent heat capacity measurement of CoNb$_4$Se$_8$. A zoom of the temperature interval where the anomaly is present is highlighted in the inset.}
		\label{figheatcap}
	\end{figure}
	
	Additionally, to confirm the nature of the magnetic transition, we performed specific heat capacity measurements on samples selected from the same growth batch as the crystals used for spectroscopic measurements, shown in Fig. \ref{figheatcap}. Specific-heat measurements were performed using the relaxation method in a \SI{14}{\T} Physical Property Measurement System (PPMS; Quantum Design). A \SI{9}{\milli\g} single crystal of CoNb$_4$Se$_8$ was mounted on the calorimeter platform using Apiezon N grease. The addenda heat capacity, including contributions from the platform and grease, was measured separately prior to the sample measurement and subtracted from the total measured heat capacity in the temperature range from \SIrange{2}{250}{\K}.
	
	The measurement exhibits a clear anomaly near the magnetic transition at $T\sim\SI{165}{\K}$\rev{, with an amplitude of $\approx$\SI[per-mode=symbol,bracket-unit-denominator = false]{5}{\J\per\mol\per\K} on a total heat capacity of $\approx$\SI[per-mode=symbol,bracket-unit-denominator = false]{280}{\J\per\mol\per\K}, per mole of CoNb$_4$Se$_8$}. These values are in good agreement with the heat capacity results reported in the previous study \cite{Regmi2024}; this thermodynamic evidence confirms that the transition is a genuine bulk antiferromagnetic ordering phenomenon. \rev{No anomaly is resolved near $T_0$, where $C_p\approx$ \SIrange[per-mode=symbol,bracket-unit-denominator = false]{75}{100}{\J\per\mol\per\K} is dominated by the lattice (13 atoms per formula unit) and an electronic contribution of \SI[per-mode=symbol,bracket-unit-denominator = false]{1}{\J\per\mol\per\K} or less would be below the resolution of the present data, whose points near $T_0$ are spaced by 2--3~K. A dedicated high-resolution heat capacity study around $T_0$ would be required to detect such a contribution.}
	
	\subsubsection*{\textbf{II. STM data pertaining to the formation of the Co surface termination, step height of CoNb$_4$Se$_8$ and susceptibility.}}
	
	\renewcommand{\thefigure}{S3}
	\begin{figure}[htb]
		\includegraphics
		{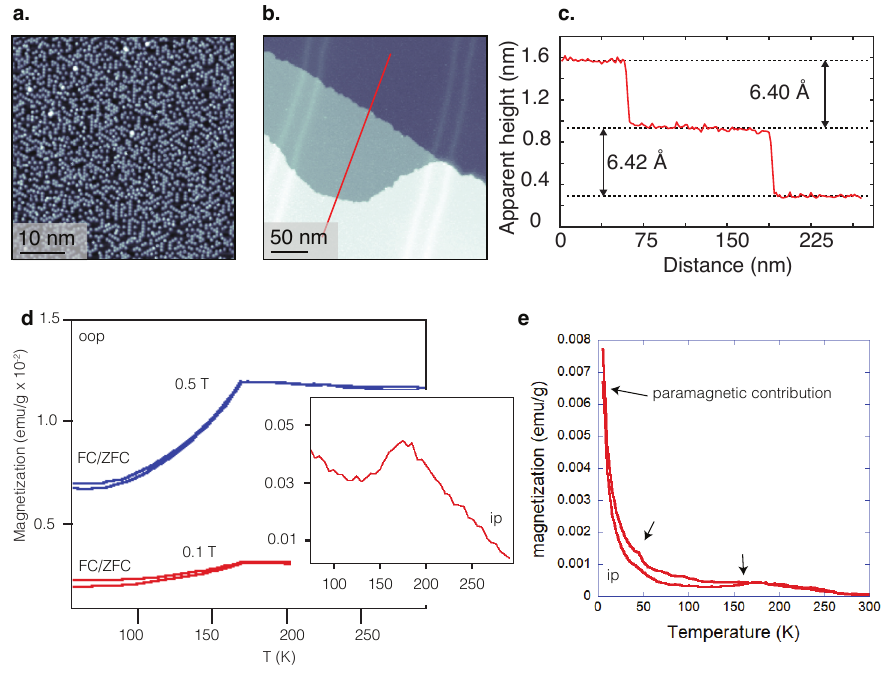}
		\caption{\textbf{STM images of the cobalt surface termination and step height analysis.} \textbf{a.} STM topographic image of the Co surface termination acquired within \SI{500}{\nano\metre} of the area shown in Fig. 1f in the main text. \textbf{b.} A large area STM topographic image showing two Se step-edges. The overlaid red line indicates where the line profile in panel c. was acquired. \textbf{c.} The line profile showing an apparent step height of $\approx\SI{6.40}{\angstrom}$, which matches closely with the $\approx\SI{6.7}{\angstrom}$ separation between Se layers within a pristine 2H-NbSe$_2$ unit cell. All images presented here were acquired with scanning parameters $V_{sample}=\SI{500}{\milli\V}$ and $I_{set}=\SI{10}{\pico\A}$. \textbf{d.} Susceptibility measurements, showing a good agreement with Ref. \cite{Regmi2024}, performed as indicated with zero field cooling and with applied field for both in plane (ip) and out of plane (oop) field. \textbf{e.} Measurements performed up to low temperature: two features - indicated by black arrows - are observed, a high-temperature one and a low-temperature one, described by both humps and change in convexity.}
		\label{figS8}
	\end{figure}
	
	In this section, we present additional STM data for the CoNb$_4$Se$_8$ sample. In Fig. \ref{figS8}a, a topographic image of the Co surface termination is shown. This image was acquired approximately \SI{500}{\nano\metre} away from the Co surface termination imaged in Fig. 1f of the main text. The two images exhibit noticeable differences: the STM image in the main text displays a well-ordered hexagonal arrangement of atoms which give rise to a 2$\times$2 modulation, whereas the image here reveals an excess of small bright protrusions with some degree of local ordering. An FFT of this STM topograph (not shown) indicates a faint 2$\times$2 modulation. While this region does not exhibit a fully developed 2$\times$2 modulation, the Co atoms appear to partially form the structure. We note that these samples were cleaved under ultra-high vacuum conditions at approximately few tens of K. Consequently, it is not surprising that the Co termination exhibits differing degrees of Co ordering.
	
	In Fig. \ref{figS8}b, a larger area of the sample is shown, displaying three distinct terraces. A line profile extracted along the red line in Fig. \ref{figS8}c reveals the apparent step height of the terraces. The measured step height of \SI{6.40}{\angstrom} closely matches the approximate \SI{6.7}{\angstrom} separation between Se layers in pristine 2H-NbSe$_2$  bulk crystal. This smaller value is consistent with the reduced $c$-axis lattice parameter measured by XRD, which indicates a step height of \SI{6.30}{\angstrom} (see Section I). In addition, we show susceptibility measurements (from SQUID) in agreement with Refs. \cite{Regmi2024, mandujano2025jacs,mandujano2025prm}: \rev{two features are observed, also evident in our optical reflectivity measurements: the magnetic transition at high temperature ($T_1\approx\SI{170}{\K}$) and a weak low-temperature anomaly ($T_0\approx\SI{50}{\K}$).} We note that the fact that the in-plane anomaly at low temperature is very small is an expected behaviour: for an easy-axis antiferromagnet (the ordered Co moments lie along the c-axis), the transverse (in-plane) susceptibility carries only a weak anomaly at $T_N$, and here it sits on the large Stoner-enhanced Pauli background of the NbSe$_2$ host. As already described in the main text, open questions remain about the exact microscopic behavior of the \rev{low-temperature anomaly}; nevertheless, this does not affect the conclusion of our work. We further note that our low-temperature feature near $T_0\approx\SI{50}{\K}$ does not coincide in temperature with the $\sim\SI{28}{}$--$\SI{36}{\K}$ transition recently assigned by Mandujano \textit{et al.} \cite{mandujano2025jacs,mandujano2025prm} to the non-centrosymmetric $\sqrt{3}\times\sqrt{3}\,R30^\circ$ ordering of Co-excess CoNb$_3$Se$_6$, and that the same authors separately describe an unresolved low-temperature anomaly in nominally stoichiometric CoNb$_4$Se$_8$ samples with which our observation is phenomenologically compatible.
	
	\subsubsection*{\textbf{III. Chemical characterization of CoNb$_4$Se$_8$}}
	
	\renewcommand{\thefigure}{S4}
	\begin{figure}[htb]
		\includegraphics{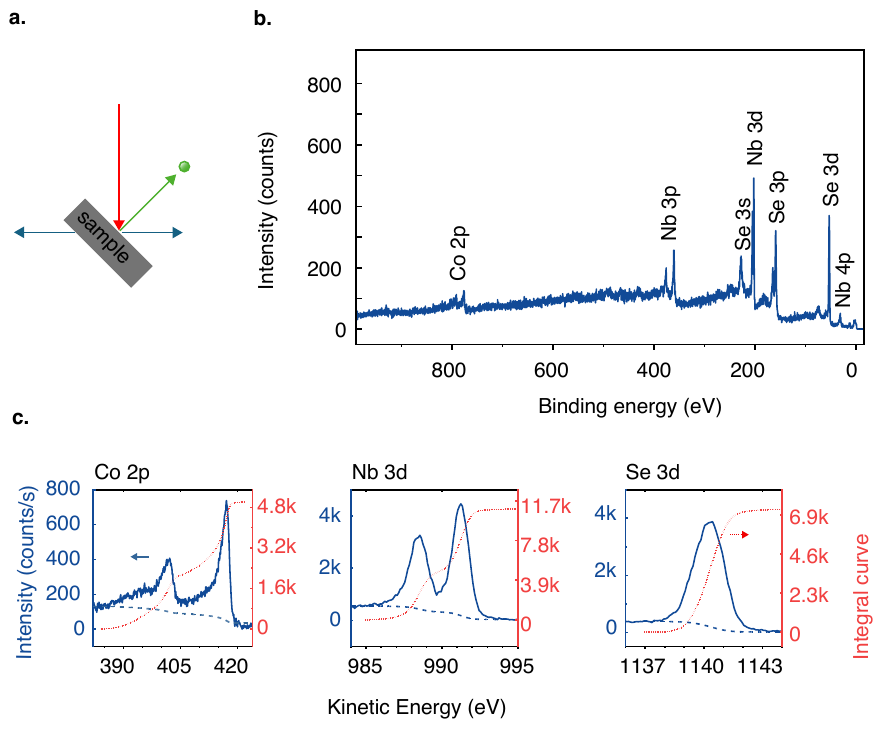}
		\caption{\textbf{XPS measurements to identify elemental contributions.} \textbf{a.} Schematic of the measurement setup, illustrating the X-ray (red arrow) and sample (gray block) geometry, the direction of the incoming electric field (double-headed blue arrow), and the emission of a photoemitted electron (green arrow and sphere). X-rays with a photon energy of \SI{1200}{\eV} were linearly polarized in the horizontal plane and incident on the sample surface at \ang{45}. \textbf{b.} Overview XPS scan displaying intensity as a function of binding energy. Dominant peaks are labeled and correspond to the expected peaks for CoNb$_4$Se$_8$. The data confirms that cleaving under ultra-high vacuum conditions ($\approx\SI{1e-10}{\milli\barpressure}$) at low temperature (\SI{70}{\K}) produces clean surfaces with no detectable contamination. \textbf{c.\textendash e.} Core-level spectra for Co 2\textit{p}, Nb 3\textit{d} and Se 3\textit{d} used to quantify the stoichiometry of the sample. The dashed blue line beneath the peaks represents the subtracted Shirley background. The core levels were integrated, with the red scale shown on the right of each panel and the as-measured intensity displayed on the left blue axes.}
		\label{figS2}
	\end{figure}
	
	The chemical composition of the sample was qualitatively and quantitatively analyzed using synchrotron-based XPS in the soft X-ray regime (APE-HE beamline at the Elettra synchrotron radiation facility in Trieste, Italy). The measurement geometry is illustrated in Fig. \ref{figS2}a. Linearly polarized X-rays with an energy of \SI{1200}{\eV} (red arrow) were incident on the sample surface at an angle of \ang{45}, with the polarization direction indicated by the blue arrow. A photoemitted electron is represented by the green circle and arrow. An overview XPS spectrum of the sample is shown in Fig. \ref{figS2}b, where the major peaks are identified and correspond to the expected elemental composition of the sample. This result confirms that the sample is free of contaminants, demonstrating that cleaving under ultrahigh vacuum conditions at low temperatures is an effective method for sample preparation. 
	
	As shown in Fig. \ref{figS2}c--d, high-resolution spectra were acquired around the Co 2\textit{p}, Nb 3\textit{d}, and Se 3\textit{d} peaks to determine the stoichiometry of the sample, confirming that it matches the expected composition. The dashed blue line beneath the data in each panel represents the subtracted Shirley background. The as-measured peak intensities are shown on the left axis in blue, while the integrated values are displayed on the right axis in red. The Co 2\textit{p} peaks are located at \SI{402}{\eV} and \SI{417}{\eV} kinetic energy, with an expected separation of \SI{15}{\eV}. The Nb 3\textit{d} spectrum exhibits a relatively larger cross-section than Co 2\textit{p}, with two well-defined peaks at \SI{988.5}{\eV} and \SI{991}{\eV} and a separation of \SI{2.5}{\eV}. The Se 3\textit{d} spectrum, also characterized by a large cross-section, displays a single dominant peak at \SI{1140}{\eV}. Together, the Nb 3\textit{d} and Se 3\textit{d} core-level spectra indicate a NbSe$_2$ chemical composition. Comparison with the Co 2\textit{p} spectrum reveals a stoichiometry of CoNb$_4$Se$_8$, with one Co atom for every four Nb atoms.
	
	\subsubsection*{\textbf{IV. Time-resolved optical reflectivity measurements of CoNb$_4$Se$_8$}}
	
	We performed time-resolved reflectivity measurements on the CoNb$_4$Se$_8$  sample at the T-ReX Laboratory of the FERMI Free-Electron Laser (FEL) at Elettra. This section is dedicated to understand the ultrafast behavior observed in pump-probe optical spectroscopy: we show here that the optical excitation employed in our experiments does not modify the underlying $2 \times 2$ arrangement of the crystal lattice.
	
	\renewcommand{\thefigure}{S5}
	\begin{figure}[ht]
		\includegraphics[width=\textwidth]{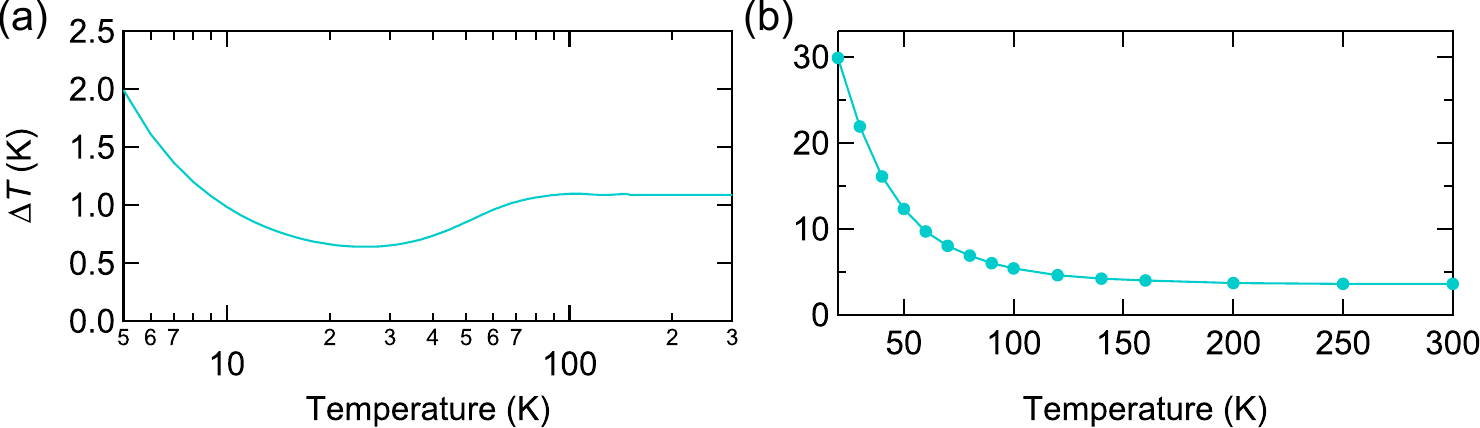}
		\caption{\textbf{Average and Phonon Heating in pump-probe.} \textbf{a.} Steady-state temperature offset resulting from the average heating of the crystal due to the average pump power. \textbf{b.} Calculated single-pulse impulsive temperature increase for the lattice phonons.}
		\label{heating}
	\end{figure}
	
	
	The possibility of a laser-induced modification of the Co $2\times2$ superstructure must be addressed. The steady-state pump-induced heating of the sample was estimated to lie below \SI{3}{\K}. This value is based on a calculation of the average crystal heating, using as inputs the deposited pump power and the temperature-dependent thermal conductivity of the material, taken from Ref. \cite{Beletskii1998}: the temperature offset $\Delta T_{AVG}$ resulting from this heating is shown in Fig.\ref{heating}a. Given the moderate value of $\Delta T_{AVG}$, no significant impact on the long-range ordering of Co atoms is expected. The electronic origin of the observed ultrafast signal is supported by a timescale analysis. In particular, the reflectivity change $\Delta R/R$ measured around \SI{650}{\nano\metre} displays a prompt onset ($<\SI{200}{\femto\s}$) following the arrival of the pump pulse, and decays rapidly. To further exclude a thermally driven lattice origin, we calculated the single-pulse impulsive lattice temperature increase $\Delta T(T)$, taking into account the lattice specific heat.
	The results are reported in \rev{Fig.~\ref{heating}b}. These calculations were performed by solving the heat equation, incorporating the temperature-dependent specific heat $C(T)$ of the lattice, following $\Delta T(T) = Q / [mC(T)]$, where $m$ is the illuminated mass and $Q$ is the absorbed energy. The high-temperature behavior of $C(T)$ was extrapolated using a Debye temperature of approximately \SI{220}{\K} \cite{Feldman1976}.

	The impulsive lattice phonons heating occurs on a timescale of \SIrange{2}{5}{\pico\s} after excitation. By contrast, the transient reflectivity signal is already suppressed by this delay time, ruling out a lattice-driven effect. This confirms that the effect observed near $t\sim\SI{100} {\femto\s}$ arises from the electronic degree of freedom. 

	\clearpage
	
	\subsubsection*{\textbf{V. Additional ARPES measurements of CoNb$_4$Se$_8$}}
	
	The data acquired using different ARPES techniques are colour-coded for clarity. ARPES data with few-micrometer-sized spatial resolution (microARPES) are presented in a ``bone" color scheme, high-energy resolution ARPES data in ``purple/white," and spin-ARPES data in ``red/blue." The low-temperature micro-ARPES measurements were conducted at the I05 beamline at the Diamond Light Source in Didcot, United Kingdom, and the room temperature data at the SGM4 beamline at ASTRID2 in Aarhus, Denmark \cite{Bianchi2023, Jones2025}. High-resolution ARPES data were collected at the CASSIOPEE beamline at Soleil Synchrotron in France, while spin-ARPES data were obtained at the LE-APE beamline at the Elettra Synchrotron in Trieste, Italy. \rev{All photon energies are converted to $k_z$ with a single set of parameters, given in Section VIII.}

	In this section, we provide additional ARPES data acquired at additional photon energies for low, intermediate and high temperatures to supplement the ARPES data in the main text. 
	
	\renewcommand{\thefigure}{S6}
	\begin{figure}[htb]
		\includegraphics{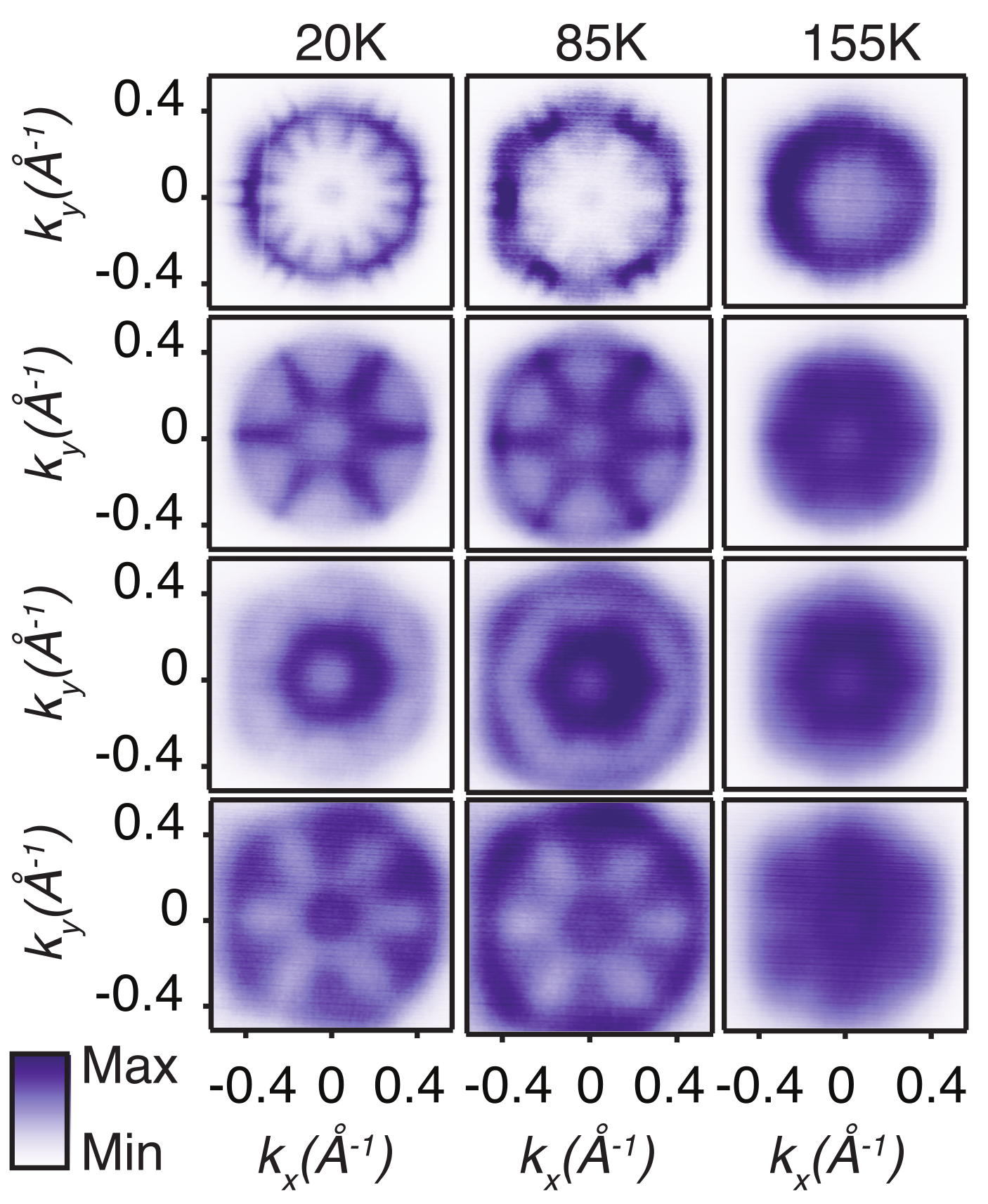}
		\caption{\textbf{CoNb$_4$Se$_8$ constant energy maps at different sample temperatures.} Constant energy maps obtained using photons with an energy of \SI{25}{\eV} are shown for $E - E_F = \SI{0}{\eV}$, $E - E_F = \SI{0.1}{\eV}$, $E - E_F = \SI{0.2}{\eV}$, and $E - E_F = \SI{0.3}{\eV}$ at three different sample temperatures: \SI{20}{\K}, \SI{85}{\K}, and \SI{155}{\K}.}
		\label{figS6}
	\end{figure}
	
	\renewcommand{\thefigure}{S7}
	\begin{figure}[htb]
		\includegraphics{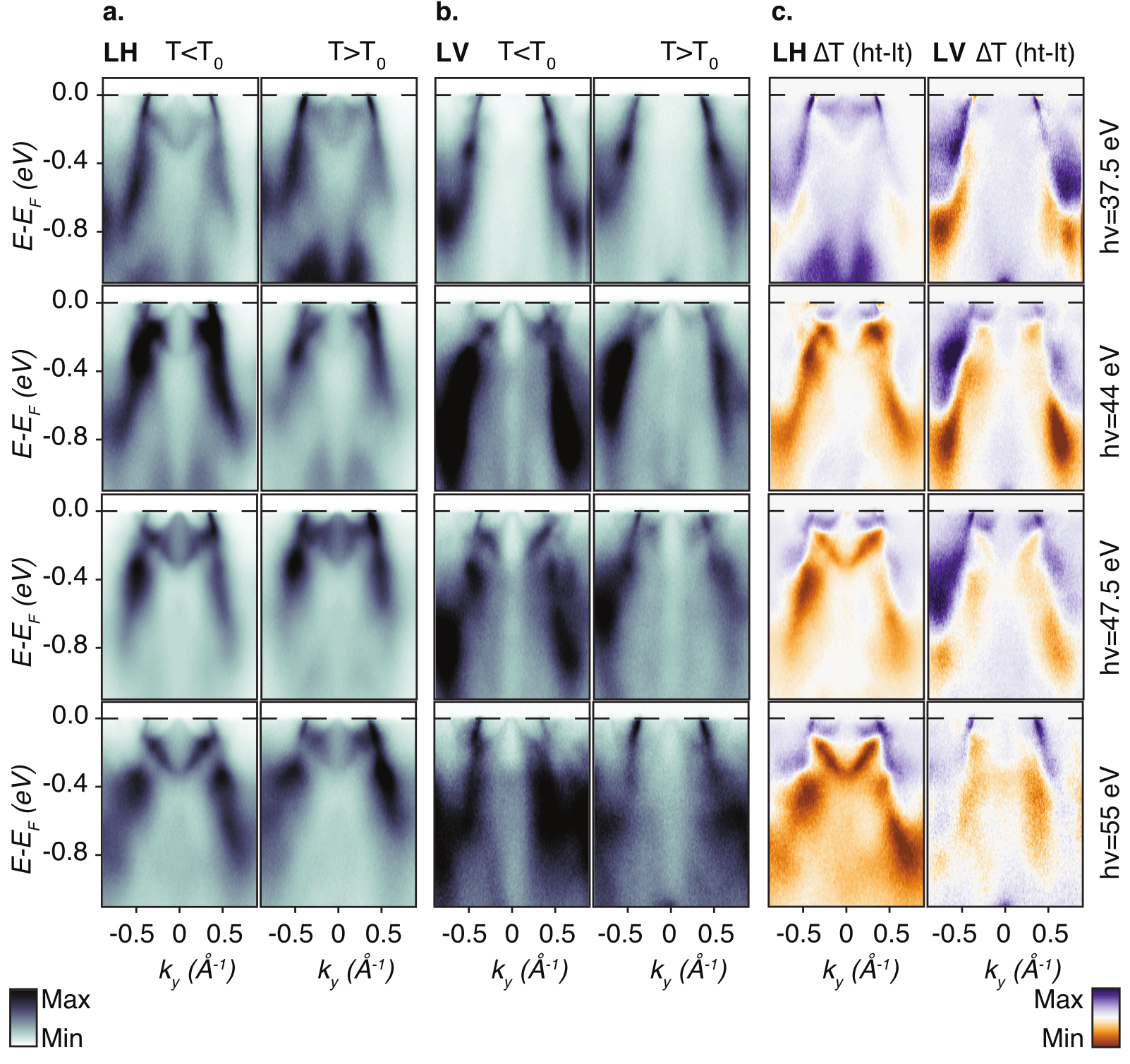}
		\caption{\textbf{MicroARPES measurements of CoNb$_4$Se$_8$ were acquired above and below $T_0$.} \textbf{a.} Comparison of the electronic structure across \rev{$T_0$}, with data collected below $T_0$ (left column) and above $T_0$ (right column) along the $\Gamma$--M direction for three photon energies, as indicated on the right-hand side. The spectra were recorded using LH polarization. \textbf{b.} Corresponding measurements obtained with LV polarization. \textbf{c.} Difference spectra highlighting changes between the high- and low-temperature electronic structures, acquired from the same spot on the sample using both polarizations. The data reveal pronounced temperature-dependent modifications, consistent with a reduction in spectral bandwidth above \rev{$T_0$}.}
		\label{figS9}
	\end{figure}
	
	\renewcommand{\thefigure}{S8}
	\begin{figure}[htb]
		\includegraphics{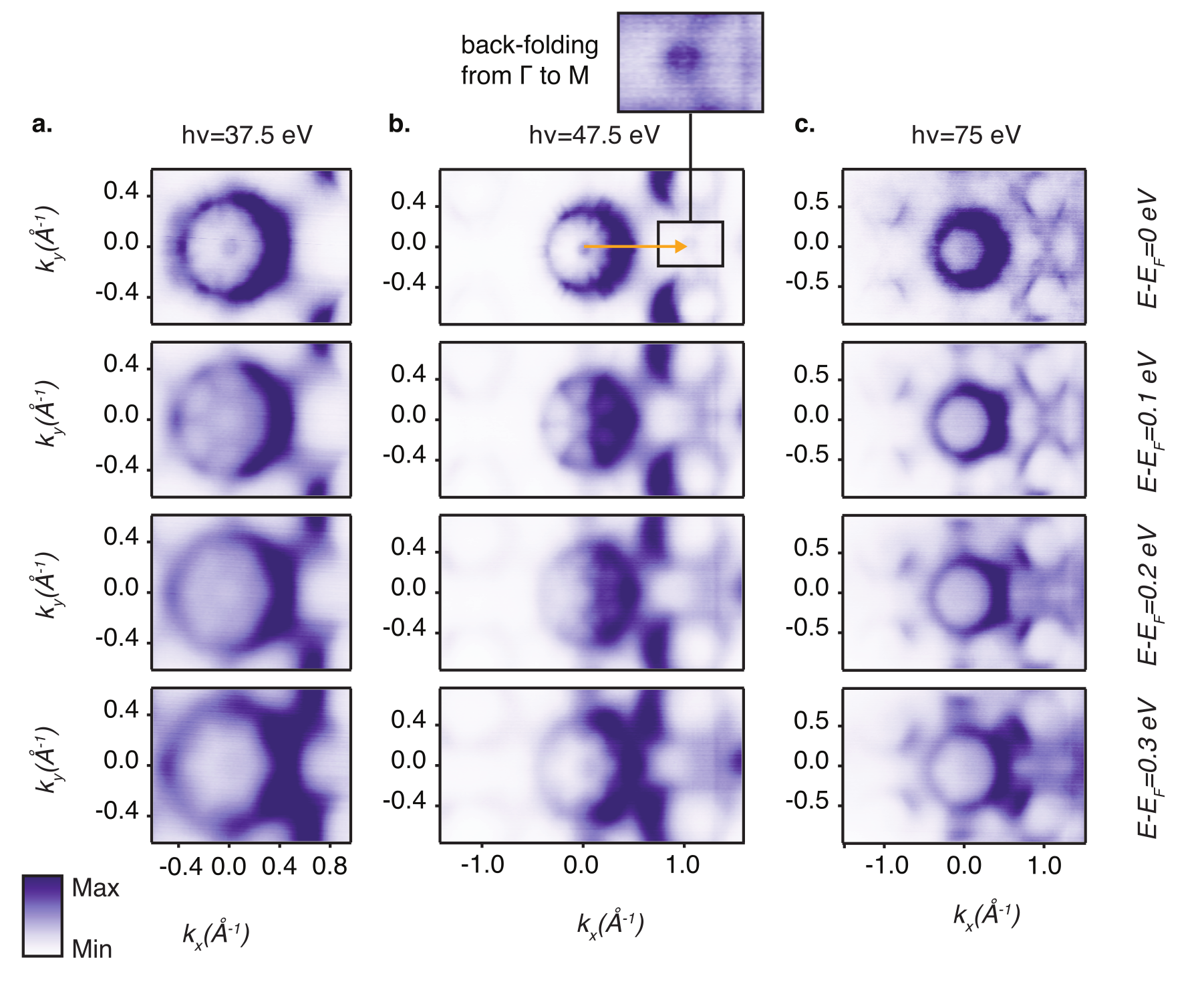}
		\caption{\textbf{Comparison of CoNb$_4$Se$_8$ constant energy maps acquired at different photon energies.} \textbf{a.} Constant energy maps obtained using photons with an energy of \SI{37.5}{\eV} are shown for $E - E_F = \rev{\SI{0}{\eV}}$, $E - E_F = \SI{0.1}{\eV}$, $E - E_F = \SI{0.2}{\eV}$, and $E - E_F = \SI{0.3}{\eV}$. This photon energy corresponds \rev{nominally to the $\Gamma$ point} in the 3D BZ \rev{(see Section VIII)}. \textbf{b.} Corresponding measurements acquired using a photon energy of \SI{47.5}{\eV} and \textbf{c.} \SI{75}{\eV}. The inset shows the back-folding of the bands from $\Gamma$ to M. \rev{The photon energies of \SI{47.5}{\eV} and \SI{75}{\eV} correspond nominally to an intermediate $k_z$ and to the A point, respectively.}}
		\label{figS7}
	\end{figure}
	
	In Fig.~\ref{figS6}, we compare constant energy maps acquired using \SI{25}{\eV} at three different temperatures: \SI{20}{\K}, \SI{85}{\K}, and \SI{155}{\K}. For each photon energy and temperature, constant energy maps are presented at \(E - E_F = \SI{0}{\eV}\), \(E - E_F = \SI{0.1}{\eV}\), \(E - E_F = \SI{0.2}{\eV}\), and \(E - E_F = \SI{0.3}{\eV}\). At first glance, the Fermi surface resembles that of bulk 2H-NbSe$_2$, with a large hexagonal pocket at the zone center, as evident in the constant energy maps acquired with \SI{70}{\eV} photons. However, notable differences are evident, where back-folding of the bands is observed, consistent with a 2$\times$2 reconstruction. As the sample temperature increases from \SI{20}{\K} to \SI{155}{\K}, the back-folding of the bands disappears, accompanied by an overall thermal broadening of the bands.
	
	\renewcommand{\thefigure}{S9}
	\begin{figure}[htb]
		\includegraphics{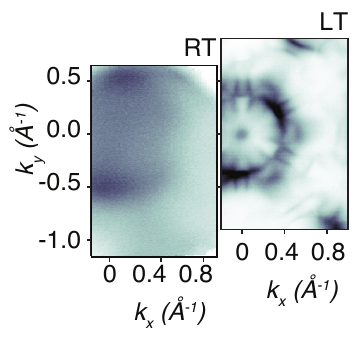}
		\caption{\textbf{Comparison of CoNb$_4$Se$_8$ constant energy maps acquired at different sample temperature.} Fermi surface ($E-E_F=0$) acquired with the sample fixed at 300\,K (left) and 25\,K (right) at a photon energy of 45.\,eV and LV polarization. }
		\label{figS9b}
	\end{figure}
	
	\renewcommand{\thefigure}{S10}
	\begin{figure}[htb]
		\includegraphics{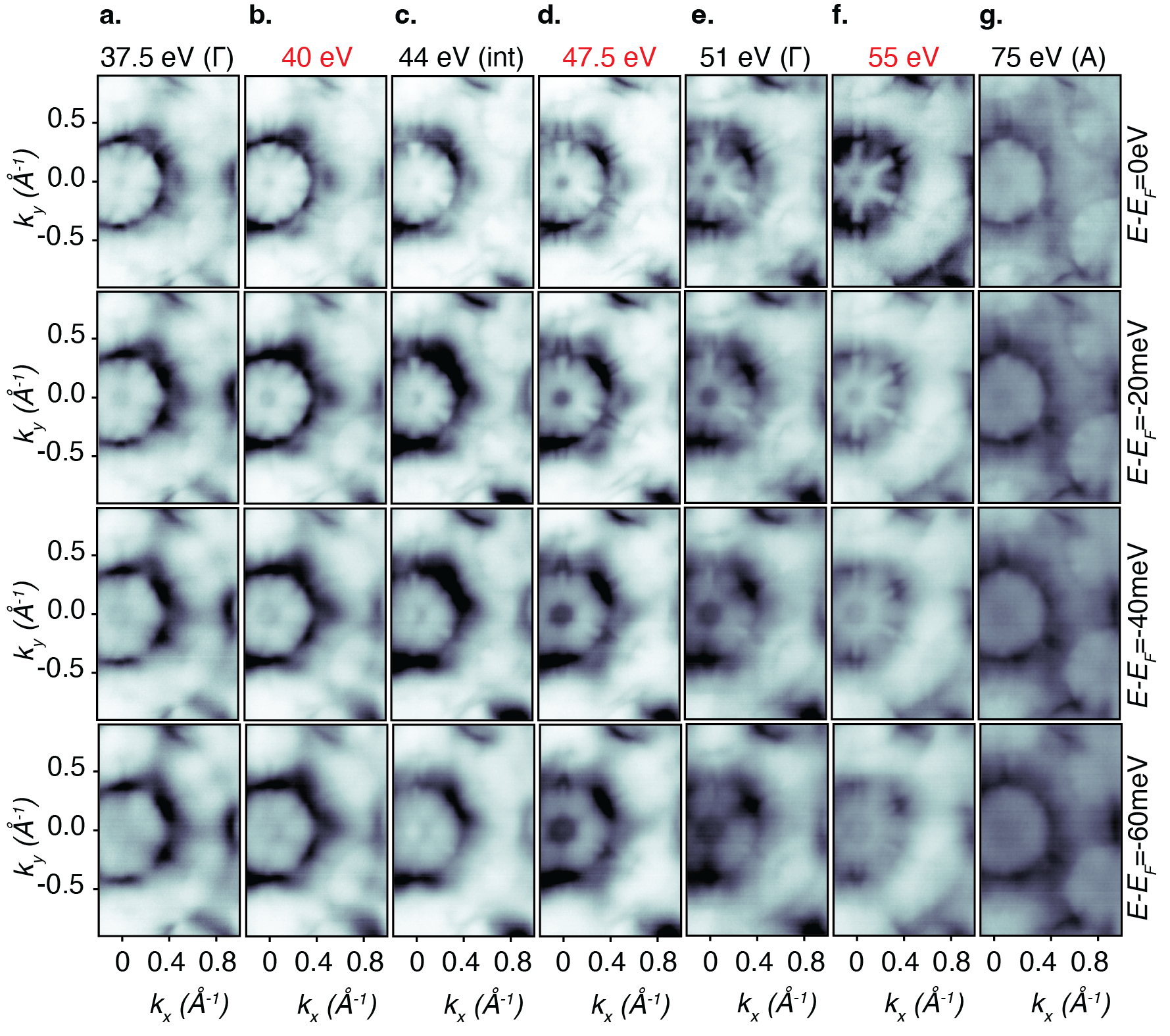}
		\caption{\textbf{CoNb$_4$Se$_8$ constant energy maps acquired with microARPES.} From \textbf{a} to \textbf{g} multiple photon energies have been used ranging across multiple BZs in $k_z$. The bulk high symmetry points have been clearly stated, with red colour (or ``int") indicating a $k_z$ in between the A-point and $\Gamma$-point. \rev{The $k_z$ values assigned to each photon energy are the nominal centres of $k_z$ windows of width $\approx\SI{0.2}{\angstrom^{-1}}$, obtained with $V_0=14\pm3$~eV and $\phi=4.5\pm0.3$~eV (see Section VIII): bulk $\Gamma$ lies near 37.5 and \SI{51}{\eV}, A near 44 and \SI{75}{\eV}, and intermediate $k_z$ near 47.5 and \SI{55}{\eV}.} Each column indicates a different binding energy to help visualize the electronic structure below the Fermi level.}
		\label{figS10}
	\end{figure}
	
	Additionally, in Fig.~\ref{figS9}, we present microARPES measurements acquired from the same sample position at different photon energies (\SI{37.5}{\eV}, \SI{44}{\eV}, \SI{47.5}{\eV}, and \SI{55}{\eV}) and light polarizations (LH and LV) at two distinct temperatures, \SI{24}{\K} and \SI{70}{\K}. The data reveal significant changes in the overall spectral bandwidth with temperature. Notably, these changes cannot be explained by a simple rigid shift model, as many spectral features remain fixed in energy and momentum. Instead, the observed temperature-dependent modifications \rev{point to a change of the electronic structure across $T_0$, whose origin remains open}. Despite the broader spectral features at higher temperatures, the energy splitting remains discernible, \rev{as expected since both temperatures lie well below $T_N$}.
	
	\renewcommand{\thefigure}{S11}
	\begin{figure}[htb]
		\includegraphics[width=0.90\columnwidth]{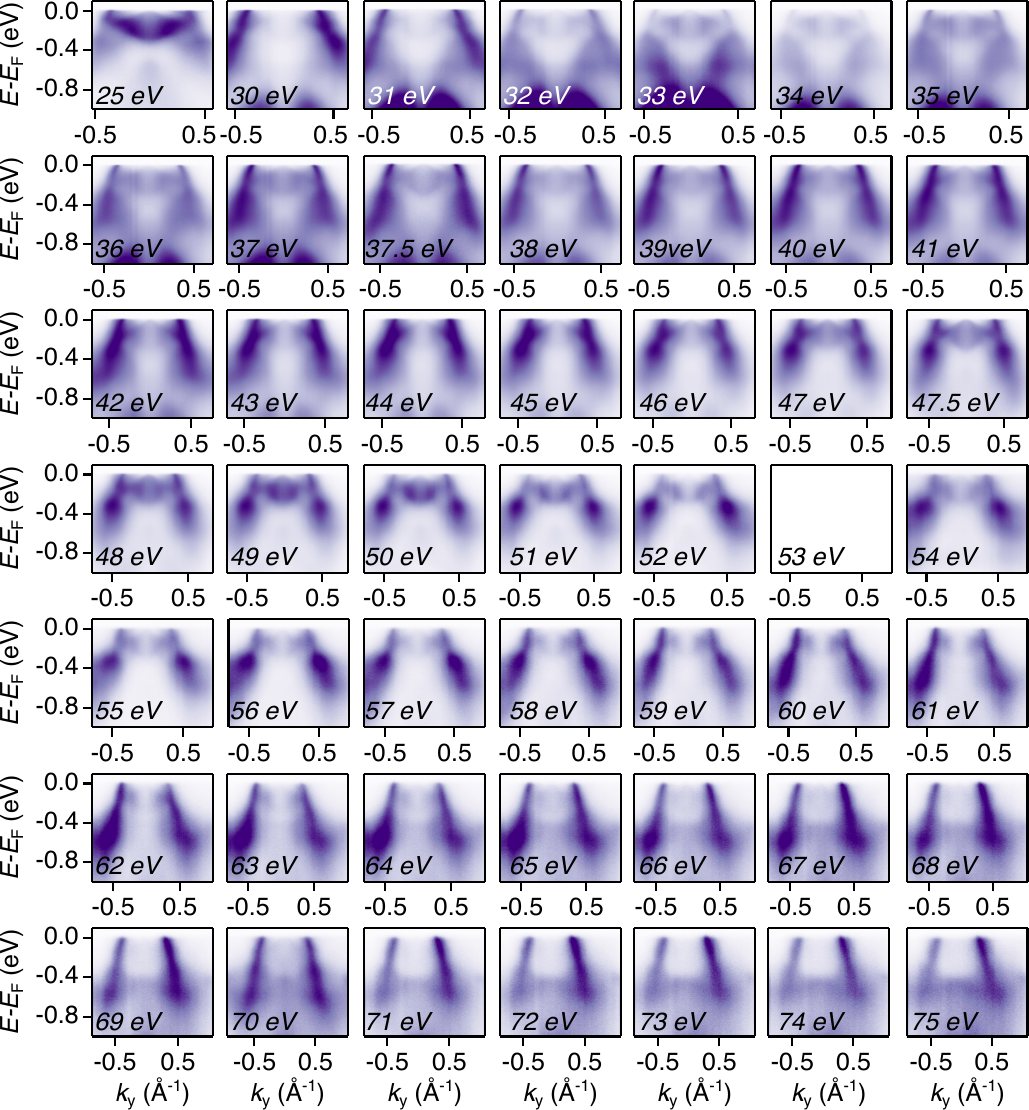}
		\caption{\textbf{CoNb$_4$Se$_8$ energy versus momentum spectra along $\Gamma$-M.} The electronic structure measured by using LH light polarization is shown as function of photon energy.}
		\label{figgm}
	\end{figure}
	
	\renewcommand{\thefigure}{S12}
	\begin{figure}[htb]
		\includegraphics[width=0.90\columnwidth]{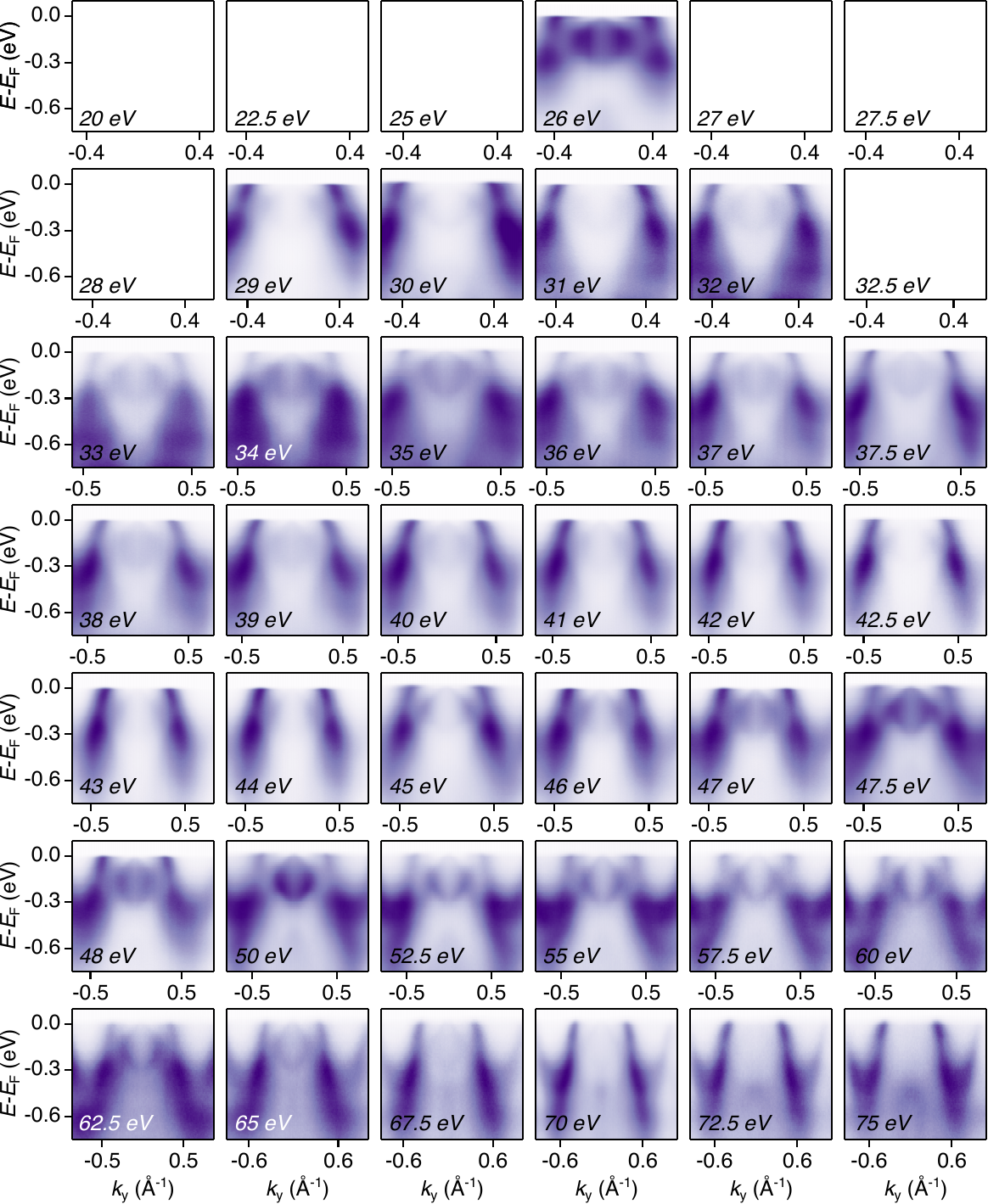}
		\caption{\textbf{CoNb$_4$Se$_8$ energy versus momentum spectra along $\Gamma$-K.} The electronic structure measured by using LH light polarization is shown as function of photon energy.}
		\label{figgk}
	\end{figure}
	
	In Fig.~\ref{figS7}, we present constant energy maps of CoNb$_4$Se$_8$ at \(E - E_F = \SI{0}{\eV}\), \(E - E_F = \SI{0.1}{\eV}\), \(E - E_F = \SI{0.2}{\eV}\), and \(E - E_F = \SI{0.3}{\eV}\) for three photon energies: \SI{37.5}{\eV}, \SI{47.5}{\eV}, and \SI{75}{\eV} \rev{(nominally $\Gamma$, intermediate $k_z$ and A, respectively)}. At the Fermi level, the fermiology predominantly exhibits a two-dimensional character. Varying the photon energy effectively tunes the matrix elements, enhancing the visibility of distinct spectral features.
	
	In Fig.~\ref{figS9b}, we compare constant-energy maps of Co\(_{1/4}\)NbSe\(_2\) at \(E - E_F = 0\) acquired using microARPES at \SI{300}{K} (left panel) and \SI{25}{K} (right panel). At \SI{300}{K}, the bands are considerably broader than at \SI{25}{K}, obscuring many finer details---particularly the \(2 \times 2\) ordering. A more detailed microARPES analysis, conducted at \SI{25}{K} with additional photon energies and spanning multiple high-symmetry points along \(k_z\), is shown in Fig.~\ref{figS10}.
	
	For completeness, we also present energy versus momentum spectra acquired along the $\Gamma$--M and $\Gamma$--K high-symmetry directions in Fig. \ref{figgm} and Fig. \ref{figgk}, respectively. The primary observation is a redistribution of spectral weight across the bands, indicating the presence of complex matrix elements and necessitating that the electronic structure of CoNb$_4$Se$_8$ be investigated using multiple photon energies. Notably, the large \textit{c}-axis parameter contributes to significant $k_z$ broadening \rev{(quantified in Section VIII)}. This implies that even with a single photon energy, a substantial portion of the Brillouin zone is probed, effectively manifesting as an apparent two-dimensionality of the bands, despite their potential bulk character.
	
	Again, for completeness, we present the calculated spin-resolved Fermi surface overlaid to the experimental data in Fig.~\ref{DFT_ARPES}, and the calculated (\textit{E}, \textit{k}) spectra with the corresponding spectral weight and spin channels along the $\Gamma$--K direction, together with other high-symmetry directions in the three-dimensional Brillouin zone, in Fig.~\ref{figtheo1}, supplementing the calculated electronic structure along the $\Gamma$--M direction shown in the main text. The agreement with the experimental results is excellent, further corroborating the findings of our study.
	
	\renewcommand{\thefigure}{S13}
	\begin{figure}[htb]
		\includegraphics[width=0.4\columnwidth]{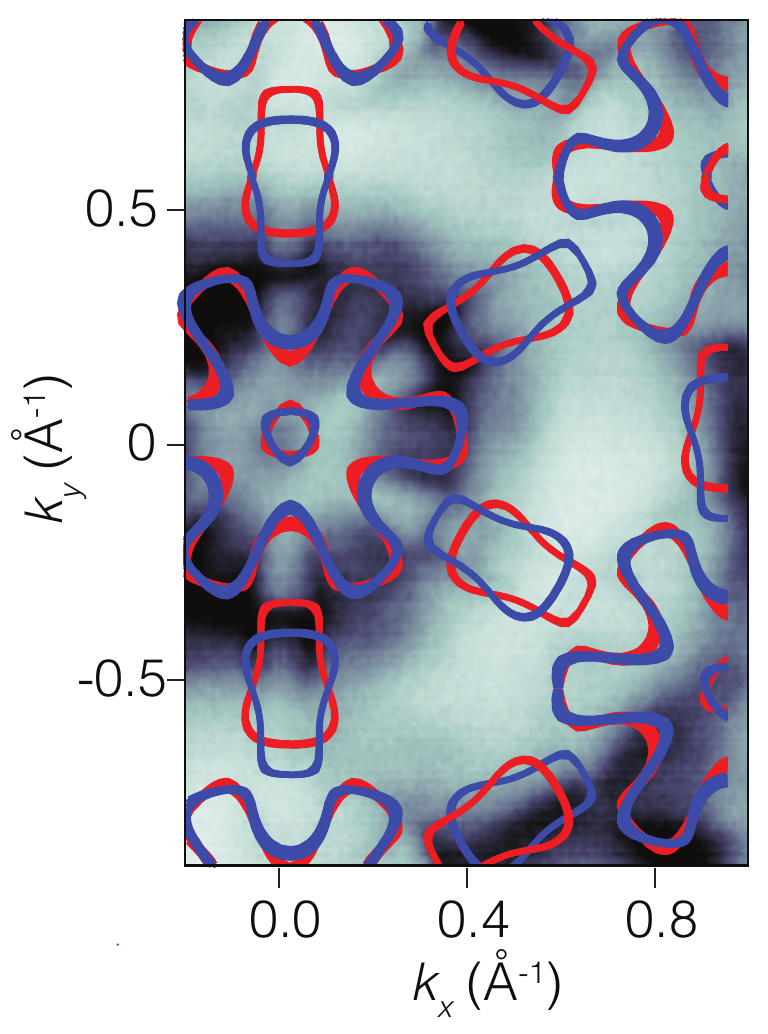}
		\caption{\textbf{ARPES Fermi surface and calculated DFT.} ARPES data have been overlaid to the DFT calculated structure. Both are considered for $k_z=0.25$; red and blue colors of the DFT indicate spin up and down respectively, and black indicates high intensity in the ARPES scale.}
		\label{DFT_ARPES}
	\end{figure}
	
	\renewcommand{\thefigure}{S14}
	\begin{figure}[htb]
		\includegraphics[width=0.9\columnwidth]{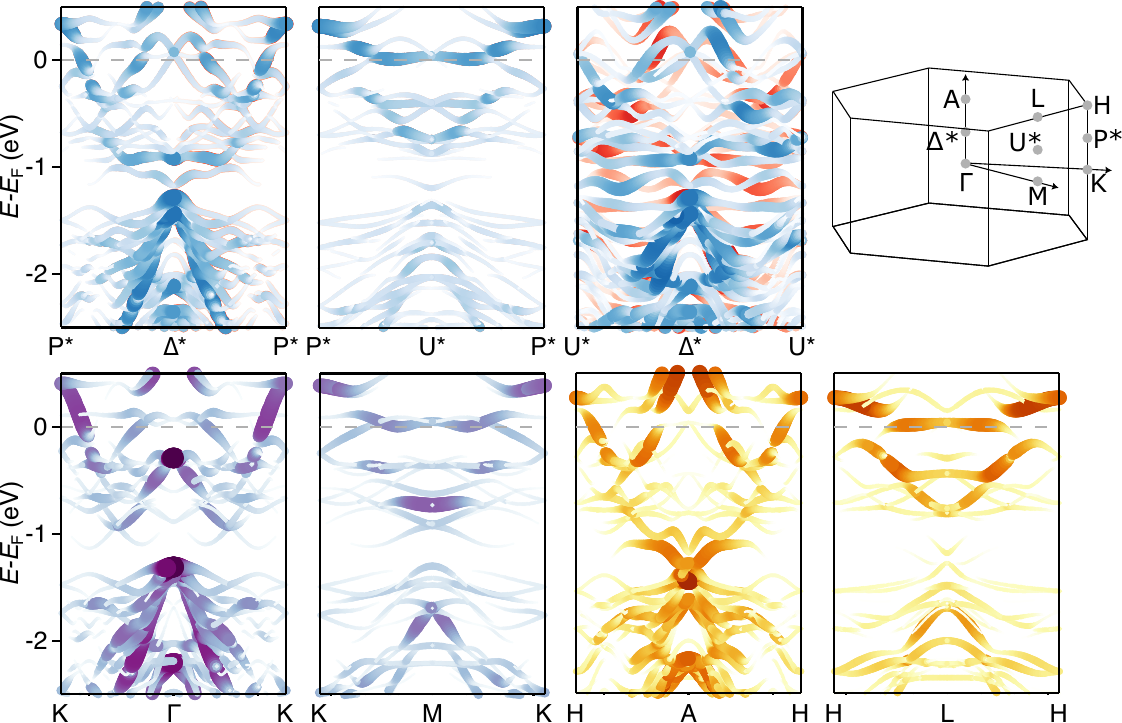}
		\caption{\textbf{CoNb$_4$Se$_8$ energy versus momentum spectra calculated by DFT along high-symmetry directions}. To complement the calculated band structure in the $\Gamma$-M direction of the main text, we present the calculated electronic structure along high-symmetry directions at different positions in $k_z$ as shown in the three-dimensional Brillouin zone.}
		\label{figtheo1}
	\end{figure}
	
	\subsubsection*{\textbf{VI. Ab-initio energetics of CoNb$_4$Se$_8$, \rev{spin space group} and the role of spin-orbit coupling}}
	
	In order to determine the most energetically favored position for the Co atoms in the $2\times2$ NbSe$_2$ supercell, we have fully relaxed the 4 possible configurations (Fig. \ref{figstruct}) and computed the total energy per Nb atom. Results are summarized in Table \ref{table:1}. We can observe that the most stable Co intercalations are such that the Co atom is directly above and below two Nb atoms.
	
	\renewcommand{\thetable}{S1}
	\begin{table}[htb]
		\centering
		\begin{tabular}{cSSS}
			\toprule
			Configuration& {\textit{a} (\si{\angstrom})} & {\textit{c} (\si{\angstrom})} & {Total energy/Nb (\si{\eV})}\\
			\midrule
			01   & 7.06    & 12.38 &   0.277\\
			02   & 7.06    & 12.40 &   0.223\\
			03   & 6.93    & 12.32 &   0.000\\
			04   & 6.95    & 12.34 &   0.043\\
			\bottomrule
		\end{tabular}
		\caption{In-plane ($a$) and out-of-plane ($c$) relaxed lattice parameters, as well as total energy per Nb atom, for each configuration of paramagnetic Co intercalated $2\times2$ NbSe2. We set to 0 the most energetically favorable configuration.} 
		\label{table:1}
	\end{table}
	
	\renewcommand{\thefigure}{S15}
	\begin{figure}[htb]
		\includegraphics[width=0.9\columnwidth]{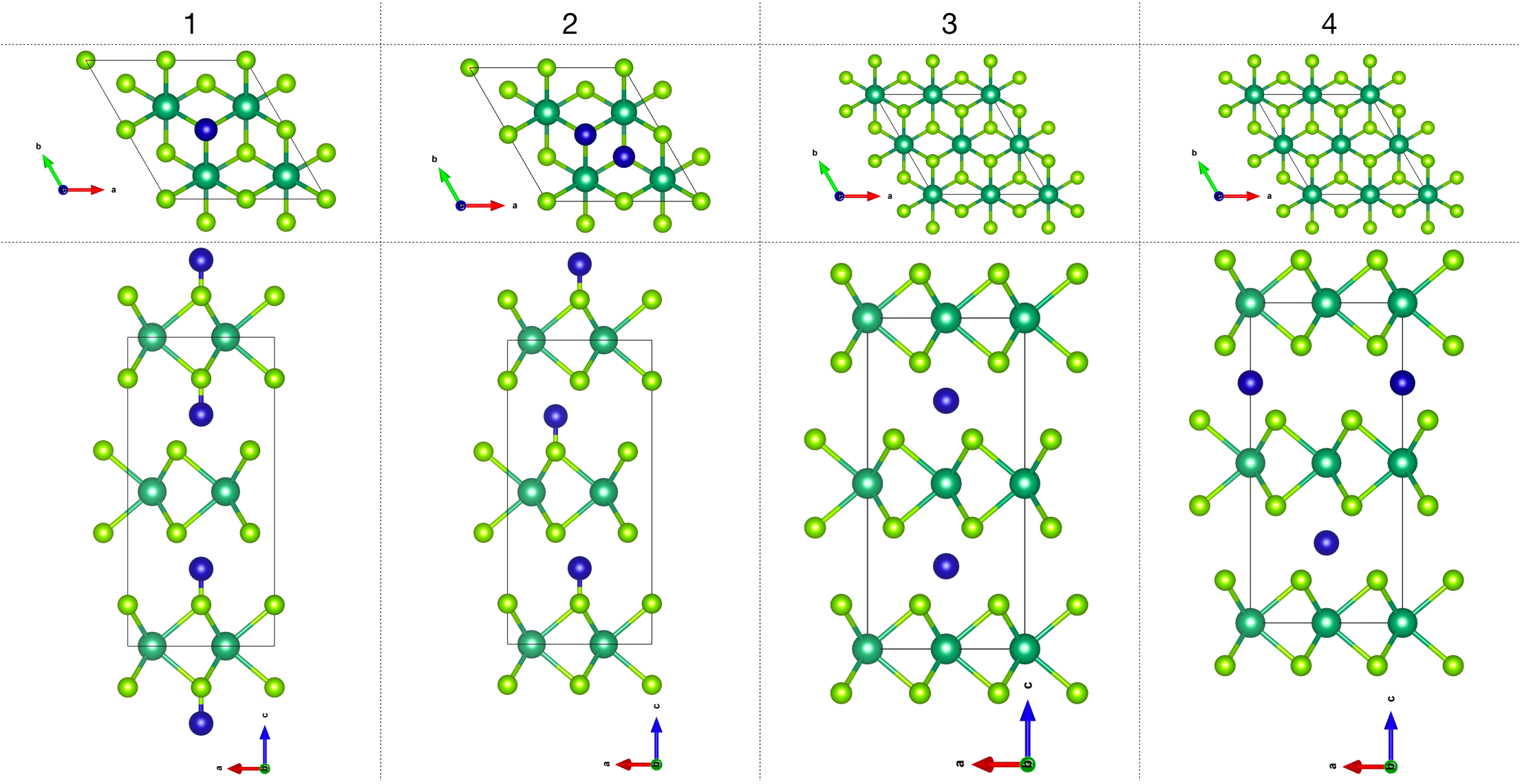}
		\caption{\textbf{Possible configurations of CoNb$_4$Se$_8$.} Stick-and-ball models (top and side views) of possible configurations of CoNb$_4$Se$_8$ analised in DFT calculations. Dark (light) green spheres are Nb (Se) atoms, while blue spheres are Co atoms.}
		\label{figstruct}
	\end{figure}
	
	\renewcommand{\thetable}{S2}
	\begin{table}[htb]
		\centering
		\begin{tabular}{ccSSS}
			\toprule
			& Configuration& {\textit{a} (\si{\angstrom})} & {\textit{c} (\si{\angstrom})} & {Total energy/Nb (\rev{\si{\milli\eV}})}\\
			\midrule
			\multirow{2}{4em}{AFM} & \textcolor{red}{03}   & \textcolor{red}{6.96}    & \textcolor{red}{12.44} &   \textcolor{red}{0.0}\\
			& 04   & 6.96    & 12.51 &   28.1\\
			\midrule
			\multirow{2}{4em}{PM} &  03  & 6.93    & 12.32 &   15.2\\
			& 04   & 6.95    & 12.34 &   58.6\\
			\midrule
			\multirow{2}{4em}{FM} & 03   & 6.94    & 12.35 &   13.9\\
			& \rev{04}   & 6.97    & 12.50 &   26.9\\
			\bottomrule
		\end{tabular}
		\caption{In-plane ($a$) and out-of-plane ($c$) relaxed lattice parameters, as well as total energy per Nb atom, for most favorable Co$_{1/4}$NbSe2 intercalations in the antiferromagnetic (AFM), paramagnetic (PM) and ferromagnetic (FM) configurations. We set to 0 the most energetically favorable \rev{AFM} configuration.}
		\label{table:2}
	\end{table}
	
	\renewcommand{\thefigure}{S16}
	\begin{figure}[htb]
		\includegraphics[width=0.9\columnwidth]{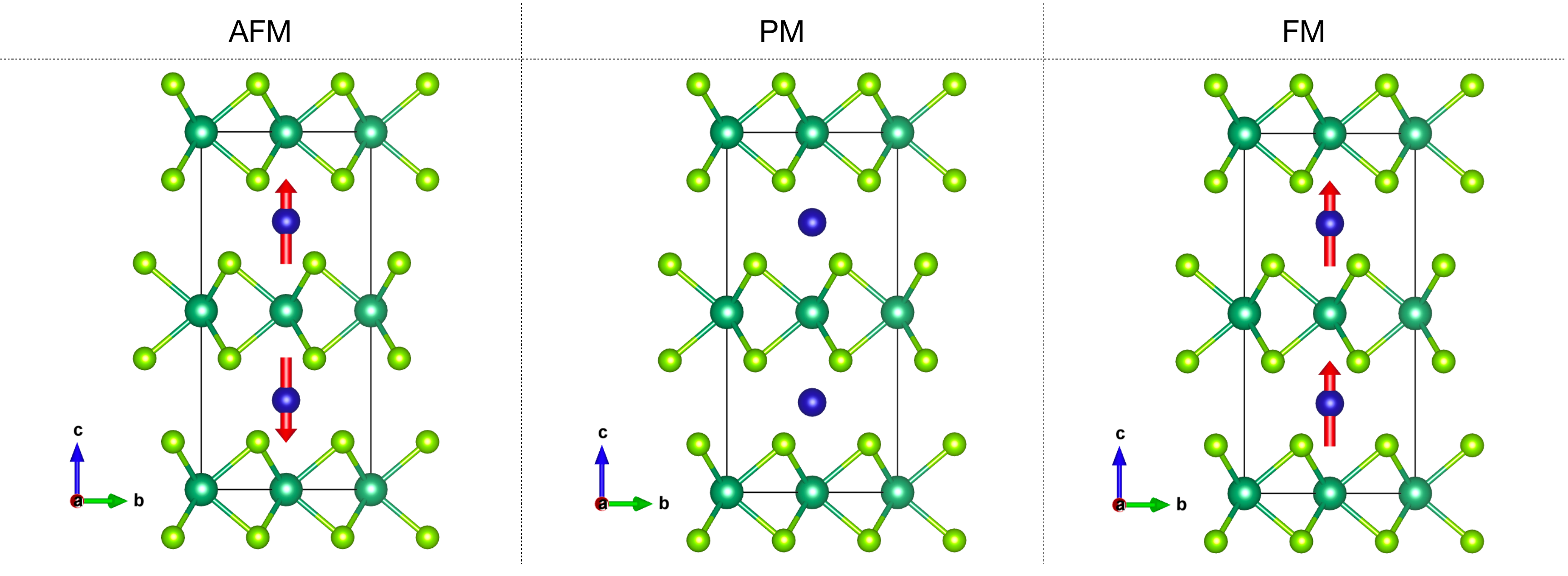}
		\caption{\textbf{Spin configurations of CoNb$_4$Se$_8$.} Stick-and-ball models of possible spin configurations of the most energetically stable CoNb$_4$Se$_8$ analised in DFT calculations. Dark (light) green spheres are Nb (Se) atoms, while blue spheres are Co atoms.}
		\label{figsmagn}
	\end{figure}
	
	After that, we proceeded to further relax the most stable configurations (03 and 04, Fig. \ref{figstruct}) taking into account three magnetic order: paramagnetic (PM, \textit{i.e.}, no net magnetization and no atomic spin), ferromagnetic (FM, net magnetization and spin aligned in the same direction) and antiferromagnetic (AFM, no net magnetization but spin aligned in opposite directions) (Fig. \ref{figsmagn}). Results are summarized in Table \ref{table:2}. We find that the system stabilizes an AFM spin configuration and that the Co atoms are aligned on top of each other. In the main text we then proceed to show electronic band structure and Fermi surfaces calculations with the AFM - 03 structure; however, since we show that Kramer degeneracy is lifted in points of the Brillouin zone which are not high-symmetry points, we call it altermagnetic (ALM).
	
	\rev{\textbf{Spin space group.} CoNb$_4$Se$_8$ crystallizes in P6$_3$/mmc (No.~194) with Co at Wyckoff position 2a: $z=0$ (sublattice A) and $z=\frac{1}{2}$ (sublattice B), with opposite moments along $c$. Each crystallographic operation is paired with the identity in spin space if it preserves the Co sublattices, and with a spin flip $C_{2\perp}$ if it exchanges them; the collinear spin-only group guarantees $E(\mathbf{k},\sigma)=E(-\mathbf{k},\sigma)$. Table~\ref{table:3} lists the resulting constraints on the nonrelativistic splitting $\Delta_s(\mathbf{k})=E_\uparrow(\mathbf{k})-E_\downarrow(\mathbf{k})$. The lowest-order harmonic compatible with them is $\Delta_s(\mathbf{k})\propto\sin(k_zc)\,k_y(3k_x^2-k_y^2)$, i.e.\ Eq.~(1) of the main text with its $k_z$ dependence written in lattice-periodic form, with nodal planes at $k_z=0$, $\pi/c$ and along the three vertical planes containing $\Gamma$--K. In Litvin notation the spin space group is P$^{-1}$6$_3$/$^{-1}$m\,$^{1}$m\,$^{-1}$c\,$^{\infty m}$1 \cite{Chen2024SSG,Xiao2024SSG,Jiang2024SSG}, the same as for the NiAs-type altermagnets MnTe and CrSb, which share the space group, the Wyckoff position of the magnetic ion and the spin arrangement \cite{Smejkal2022_1,Smejkal2022_2,Krempasky2024,Reimers_2024}. The corresponding spin point group, $^{2}6/^{2}m^{2}m^{1}m$, is given for this compound in Ref.~\cite{Dale_2024}, where the sublattice-exchanging glides are also identified as those with normals along $\langle1\bar{1}0\rangle$, as in Table~\ref{table:3}; the two symbols describe the same group. The assignment is consistent with the magnetic space group P6$_3'$/m$'$m$'$c determined by neutron diffraction \cite{Regmi2024}, with DFT (degeneracy at $k_z=0$, $\pi/c$ and along $\Gamma$--K, splitting along $\Gamma$--M at intermediate $k_z$; Fig.~2h of the main text and Fig.~\ref{figtheo1}), and with spin-ARPES (reversal along $\Gamma$--M, null along $\Gamma$--K; Section VII). This classification applies to the $x=1/4$ phase of Co$_x$NbSe$_2$ ($2\times2$ superstructure, P6$_3$/mmc). The $x=1/3$ phase ($\sqrt{3}\times\sqrt{3}$ superstructure, non-centrosymmetric P6$_3$22) has a different space group, and its spins are canted by the Dzyaloshinskii--Moriya interaction \cite{mandujano2025jacs,mandujano2025prm,Dale_2024}, so it does not belong to the same spin space group.}
	
	\renewcommand{\thetable}{S3}
	\begin{table}[htb]
		\centering
		\footnotesize
		\setlength{\tabcolsep}{4pt}
		\begin{tabular}{lccll}
			\toprule
			Operations (Seitz) & Co sublattices & Spin part & Constraint on $\Delta_s(\mathbf{k})$ & Consequence in $\mathbf{k}$ space\\
			\midrule
			\makecell{$\{1|0\}$, $\{3^{\pm}_{001}|0\}$, $\{\bar{1}|0\}$, \\ $\{\bar{3}^{\pm}_{001}|0\}$} & preserved & $E$ & \makecell{$\Delta_s(R\mathbf{k})=\Delta_s(\mathbf{k})$; \\ $\Delta_s(-\mathbf{k})=\Delta_s(\mathbf{k})$} & \makecell{threefold symmetric, \\ even in $\mathbf{k}$}\\
			\makecell{mirrors $\perp\langle100\rangle$; twofold \\ axes $\parallel\langle100\rangle$} & preserved & $E$ & \makecell{$\Delta_s$ even across\\ these planes} & \makecell{no nodal plane: splitting \\ allowed along $\Gamma$--M}\\
			$\{6^{\pm}_{001}|00\frac{1}{2}\}$, $\{\bar{6}^{\pm}_{001}|00\frac{1}{2}\}$ & exchanged & $C_{2\perp}$ & $\Delta_s(C_6\mathbf{k})=-\Delta_s(\mathbf{k})$ & \makecell{sign alternates \\every $60^\circ$}\\
			\makecell{$\{m_{001}|00\frac{1}{2}\}$ (mirror at \\ $z=\frac{1}{4}$)} & exchanged & $C_{2\perp}$ & \makecell{$\Delta_s(k_x,k_y,-k_z)=$ \\ $-\Delta_s(\mathbf{k})$} & \makecell{nodal planes $k_z=0$ \\and $k_z=\pi/c$}\\
			\makecell{$c$ glides $\perp\langle1\bar{1}0\rangle$; twofold \\ axes $\parallel\langle1\bar{1}0\rangle$ (at $z=\frac{1}{4}$)} & exchanged & $C_{2\perp}$ & \makecell{$\Delta_s$ odd across\\ these planes} & \makecell{nodal vertical planes \\containing $\Gamma$--K}\\
			\bottomrule
		\end{tabular}
		\caption{\rev{Spin space group operations of A-type CoNb$_4$Se$_8$ and their constraints on the nonrelativistic spin splitting} ($k_x\parallel\Gamma$--K, $k_y\parallel\Gamma$--M).}
		\label{table:3}
	\end{table}
	
	We have additionally performed magnetic non-collinear spin-orbit coupling (SOC) DFT calculations. In order to do so we employed full relativistic pseudopotentials and the PBE exchange-correlation functional. More precisely, we have employed ultrasoft pseudopotentials for Nb and Co, while we have used a norm-conserving pseudopotential for Se. We have increased the kinetic energy cutoff for the wavefunctions to $60$ Ry, while we kept the density cutoff to $450$ Ry. The Brillouin zone has been sampled with a Gamma centered uniform \textit{k}-point mesh of $9\times9\times16$ and a Methfessel-Paxton electronic smearing of $5$ mRy. Calculations are always performed without a Hubbard (DFT+\textit{U}) correction. Convergence hase been achieved with a threshold of $10^{-5}$ Ry by starting with and AFM spin configuration on Co atoms and by constraining the total magnetization adding a penalty functional to the total energy:
	
	\begin{equation}
		\lambda * \sum_{i} (M_{i} - M_{f,{i}})^2,
	\end{equation}
	
	where lambda = $0.01$ Ry is a Lagrange multiplier, the sum i is over the three components of the magnetization $M_{i}$ and $M_{f,{i}}$ is the fixed magnetization. 
	
	\renewcommand{\thefigure}{S17}
	\begin{figure}[htb]
		\includegraphics[width=0.99\columnwidth]{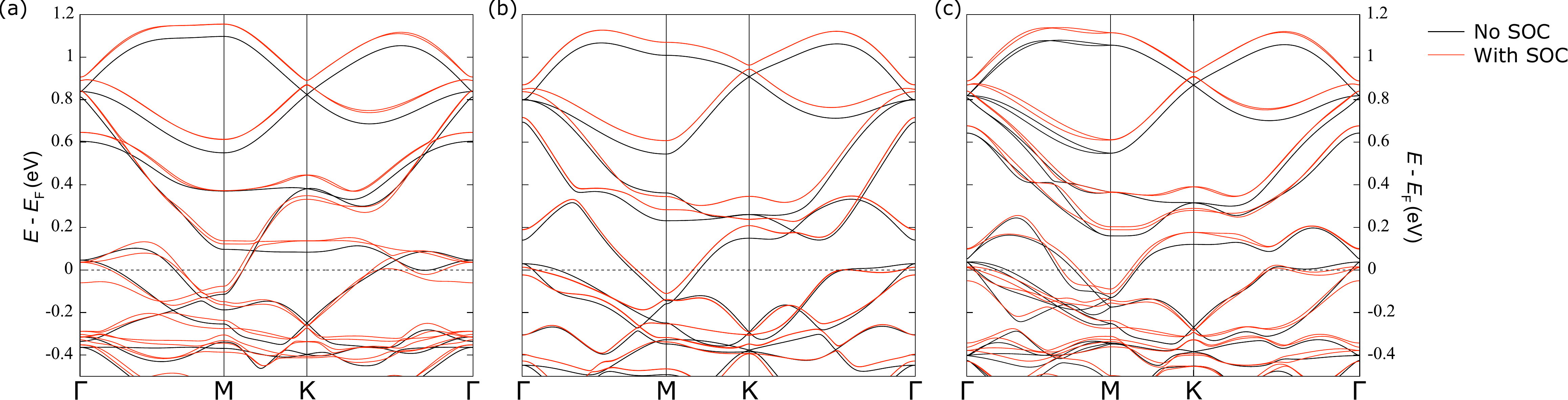}
		\caption{\textbf{The role of SOC}. Comparison between band structure calculations performed without (black lines) and with (red lines) SOC along the $\Gamma$-M direction, at $k_z=$ (a) $0$, (b) $\pi/c$, and (c) $\pi/2c$.}
		\label{SOC}
	\end{figure}
	
	The final spin configuration is an AFM orientation of the spins on the Co atoms, without any sizeable canting. In Fig.~\ref{SOC} we show the comparison of the electronic band structure without (black line) and with (red line) SOC in the small Brillouin zone of the $2\times2$ unit cell at different values of $k_z$ ( (a) $0$, (b) $\pi/c$, and (c) $\pi/2c$). We can observe that the effect of the spin-orbit coupling is mainly that of lifting band degeneracies at band crossing, with the strongest energy of $\Delta_{SOC} = \SI{0.1}{\eV}$ at $k = (0,0,0)$ and $\Delta_{SOC} = \SI{0.2}{\eV}$ at $k = (0,0,\pi/c)$. \rev{These values are SOC gaps at band crossings on the $k_z$ axis, where the altermagnetic splitting vanishes identically by symmetry (Table~\ref{table:3}); they are not spin splittings, and the SOC gaps and the altermagnetic splitting occur in different regions of the Brillouin zone.}
	
	\renewcommand{\thefigure}{S18}
	\begin{figure}[htb]
		\includegraphics[width=0.9\columnwidth]{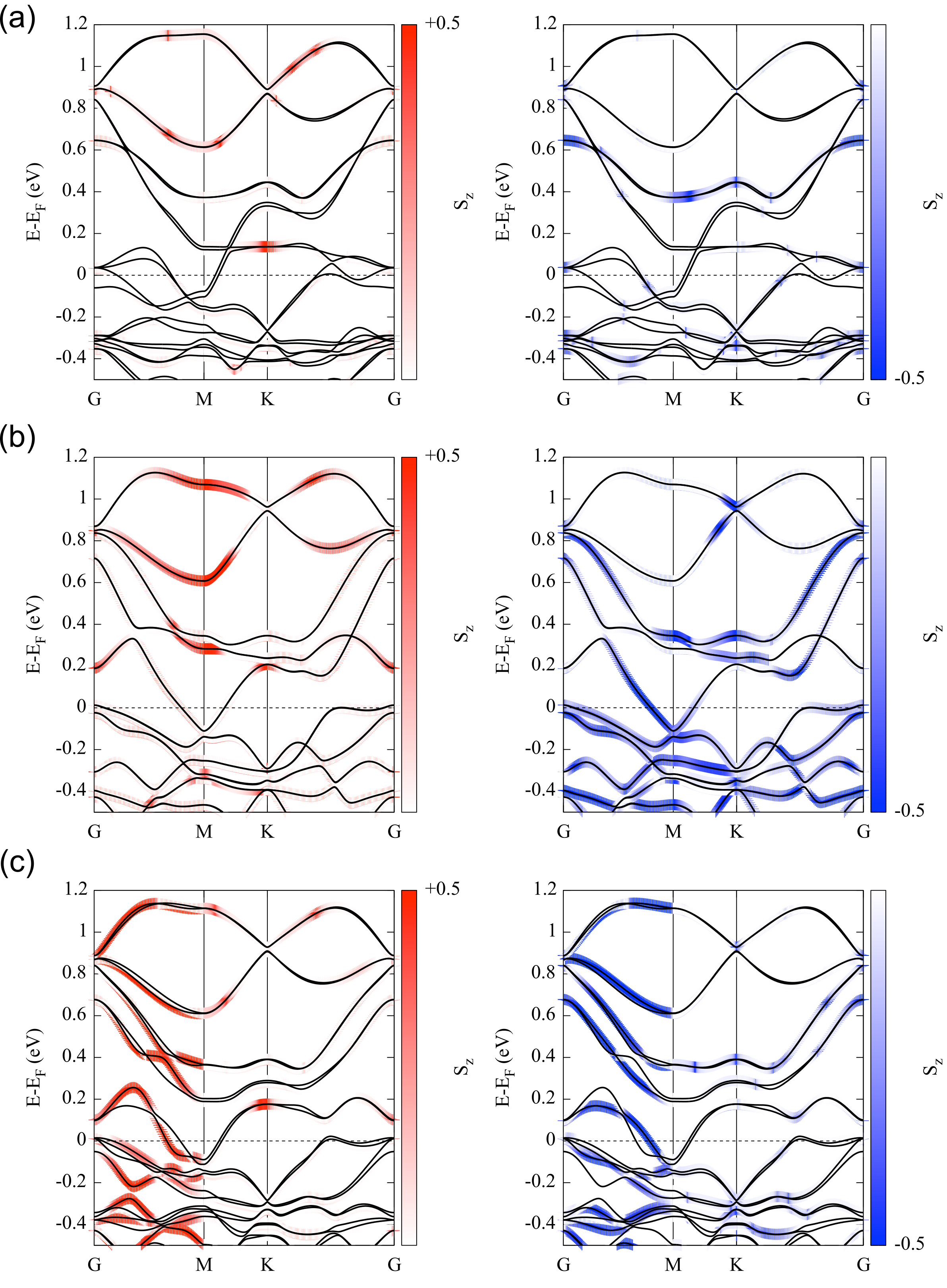}
		\caption{\textbf{Spin character of bands with SOC}. Band structure calculations with SOC along the $\Gamma$-M direction, at $k_z=$ (a) $0$, (b) $\pi/c$, and (c) $\pi/2c$. The color scale represents the projection along the \textit{z} axis of the spinor wavefunction \textit{S}, where $S_z = +0.5$ $(-0.5)$ means that the spinor is fully aligned along the positive (negative) direction of the \textit{z} axis.}
		\label{SOCspin}
	\end{figure}
	
	In order to understand if the SOC also affects the spin texture of the electronic bands, we have computed the expectation value of the spin operator on the spinor wavefunction along the out-of-plane direction, \textit{i.e.}, $S_z$. In Fig.~\ref{SOCspin} we show the electronic bands with SOC, on top of which we plotted the $S_z$ value of the eigenvalues. As we can see, SOC does not affect the spin texture that we observe without it and, more precisely, the altermagnetic splitting at $k_z = \pi/2c$ is still present along the $\Gamma$-M direction. \rev{The comparison at the momenta of the altermagnetic pairs $\pm k_{s1,s2}$ near $E_F$ at $k_z=\pi/2c$ is summarized in Table~\ref{table:4}: the nonrelativistic splitting is $\approx\SI{150}{\milli\eV}$, whereas SOC shifts these bands by less than \SI{20}{\milli\eV} and leaves $S_z\approx\pm0.5$, i.e.\ fully polarized in the convention of Fig.~\ref{SOCspin}.}
	
	\renewcommand{\thetable}{S4}
	\begin{table}[htb]
		\centering
		\setlength{\tabcolsep}{8pt}
		\begin{tabular}{ll}
			\toprule
			{Quantity} & {Value}\\
			\midrule
			{Nonrelativistic splitting of the $\pm k_{s1,s2}$ branches ($k_z=\pi/2c$, near $E_F$)} & $\approx\SI{150}{\milli\eV}$\\
			SOC-induced shift of the same bands & $<\SI{20}{\milli\eV}$\\
			$S_z$ of the same bands with SOC & $\approx\pm0.5$ (fully polarized)\\
			Largest SOC gap at a band crossing, $\mathbf{k}=(0,0,0)$ & \SI{0.1}{\eV} ($\Delta_s=0$ by symmetry)\\
			Largest SOC gap at a band crossing, $\mathbf{k}=(0,0,\pi/c)$ & \SI{0.2}{\eV} ($\Delta_s=0$ by symmetry)\\
			\bottomrule
		\end{tabular}
		\caption{\rev{Nonrelativistic splitting and SOC effects in CoNb$_4$Se$_8$}, from the calculations of Figs.~\ref{SOC} and \ref{SOCspin}.}
		\label{table:4}
	\end{table}
	
	\clearpage
	
	\subsubsection*{\textbf{VII. \rev{Spin-ARPES data treatment and} time-reversal domains in CoNb$_4$Se$_8$}}
	
	\rev{\textbf{Data treatment.} Each spin-resolved EDC was treated with the standard procedure for exchange-scattering (V-LEED) spin detectors \cite{Kessler1985,Dil2009,Okuda2011,Bigi2017}. (i) At each momentum the EDC is recorded twice, with the Fe(001)-O target of the V-LEED detector magnetized in opposite directions, giving $I_+^{\mathrm{raw}}(E)$ and $I_-^{\mathrm{raw}}(E)$. (ii) The two magnetization states of the target have slightly different reflectivity, so that the two channels have different overall efficiencies. This instrumental factor is removed by scaling the channels so that they coincide in an energy window $W$ in the high binding energy tail, dominated by inelastically scattered electrons: $I_-(E)=\alpha\,I_-^{\mathrm{raw}}(E)$, with $\alpha=\sum_W I_+^{\mathrm{raw}}(E)/\sum_W I_-^{\mathrm{raw}}(E)$. In a compensated magnet this background is expected to be unpolarized, since inelastically scattered electrons originate from a broad range of momenta, over which the altermagnetic polarization averages out; this assumption is tested independently by the checks in (vi). (iii) The asymmetry and the polarization are $A(E)=[I_+(E)-I_-(E)]/[I_+(E)+I_-(E)]$ and $P(E)=A(E)/S$, where $S=0.30$ is the effective Sherman function of the detector \cite{Bigi2017}, recalibrated routinely after each preparation of the target. (iv) The spin-resolved spectra are $I_{\uparrow,\downarrow}(E)=\frac{1}{2}I_{\mathrm{tot}}(E)\,[1\pm P(E)]$, with $I_{\mathrm{tot}}=I_++I_-$. (v) The counts follow Poisson statistics, so that $\sigma_P(E)=S^{-1}\{[1-A(E)^2]/[N_+(E)+N_-(E)]\}^{1/2}$, where $N_\pm$ are the raw counts in each energy channel; the uncertainty of $\alpha$ is negligible because $W$ contains $\approx10^5$ counts, and the uncertainty of $S$ only rescales $P$ and cannot change its sign. (vi) A residual instrumental asymmetry would be the same at $+k$ and $-k$ and on different spots of the sample, whereas the intrinsic polarization reverses in both cases; the reversal between opposite momenta and between spots therefore cannot originate from the normalization, and the same procedure gives $P=0$ within $1\sigma$ along $\Gamma$--K (Fig.~\ref{spinsi}b), confirming that it creates no artificial signal.}
	
	The presence of domains is a crucial consideration in photoemission experiments where the signal is averaged over a finite beam spot. In the case of CoNb$_4$Se$_8$, evidence for domain structure arises from the comparison between conventional ARPES (beam diameter $\sim \SIrange{30}{80}{\micro\metre}$) and micro-ARPES ($\sim \SI{5}{\micro\metre}$ spatial resolution). Notably, the former exhibits greater band broadening, consistent with averaging over regions that are not electronically homogeneous—an indication of domain formation. This consideration is particularly important in spin-resolved ARPES (spin-ARPES), where the spot size ($\sim \SIrange{30}{80}{\micro\metre}$ on average) likely encompasses multiple domains. Although this precludes the extraction of the absolute value of spin polarization in percentage, the detection of a finite spin signal and a clear reversal at opposite momenta demonstrates that the domains must be at least comparable in size to the beam footprint. If domains were significantly smaller or randomly oriented within the illuminated area, the spin contributions would statistically cancel, yielding zero polarization. This is not observed in our data. 
	
	Additionally, to further scrutinize the presence of rotated crystalline domains, spin-ARPES measurements were also performed along the $\Gamma$–K direction, where theoretical models predict zero spin polarization due to restored Kramers degeneracy. In agreement with these predictions, no spin signal is observed along this direction. This rules out significant crystalline domain rotation or misalignment, which could lead to spectral contributions from other high-symmetry directions. This null measurement is also a direct consequence of the $g$-wave form factor discussed above: the vertical mirror planes containing the $\Gamma$--K high-symmetry lines are symmetry-protected nodal planes of $\Delta_s(\mathbf{k})$, so the spin signal along $\Gamma$--K is enforced to vanish by the spin space group. \rev{In the bulk this holds even when SOC is included: the $c$ glide whose plane contains $\Gamma$--K and the $c$ axis is a symmetry of P6$_3'$/m$'$m$'$c without time reversal, and it leaves $\mathbf{k}$ invariant while reversing $\sigma_z$. At the surface this glide, which involves a $c/2$ translation, is lost, and a hidden Ising polarization of the topmost NbSe$_2$ layer becomes symmetry-allowed. Such layer-locked polarization, however, requires the states to be localized on individual layers, i.e.\ an interlayer coupling weak compared with the spin--orbit coupling, as in van der Waals crystals \cite{Zhang2014hidden,Riley2014}. In CoNb$_4$Se$_8$ adjacent NbSe$_2$ layers are covalently bonded through the intercalated Co, which delocalizes the states across the layers and is expected to strongly reduce this polarization; a spin splitting near K has been attributed to this mechanism in Ref.~\cite{Dale_2024}. Conversely, along $\Gamma$--M any time-reversal-even $\sigma_z$ (hidden Ising polarization, surface Rashba coupling, or spin polarization induced by the photoemission process for photons incident in this plane) is forbidden, since the vertical mirror containing $\Gamma$--M survives at the surface and reverses $\sigma_z$; only the magnetic order, for which this mirror is combined with time reversal (m$'$ in P6$_3'$/m$'$m$'$c), allows it (Section IX).} These observations support the conclusion that, even in the presence of domains, their orientations are predominantly aligned with the crystal symmetry axes, as is typical for high-quality single crystals. This is shown in Fig.\ref{Domain1}.
	
	\renewcommand{\thefigure}{S19}
	\begin{figure}[htb]
		\includegraphics[width=0.6\columnwidth]{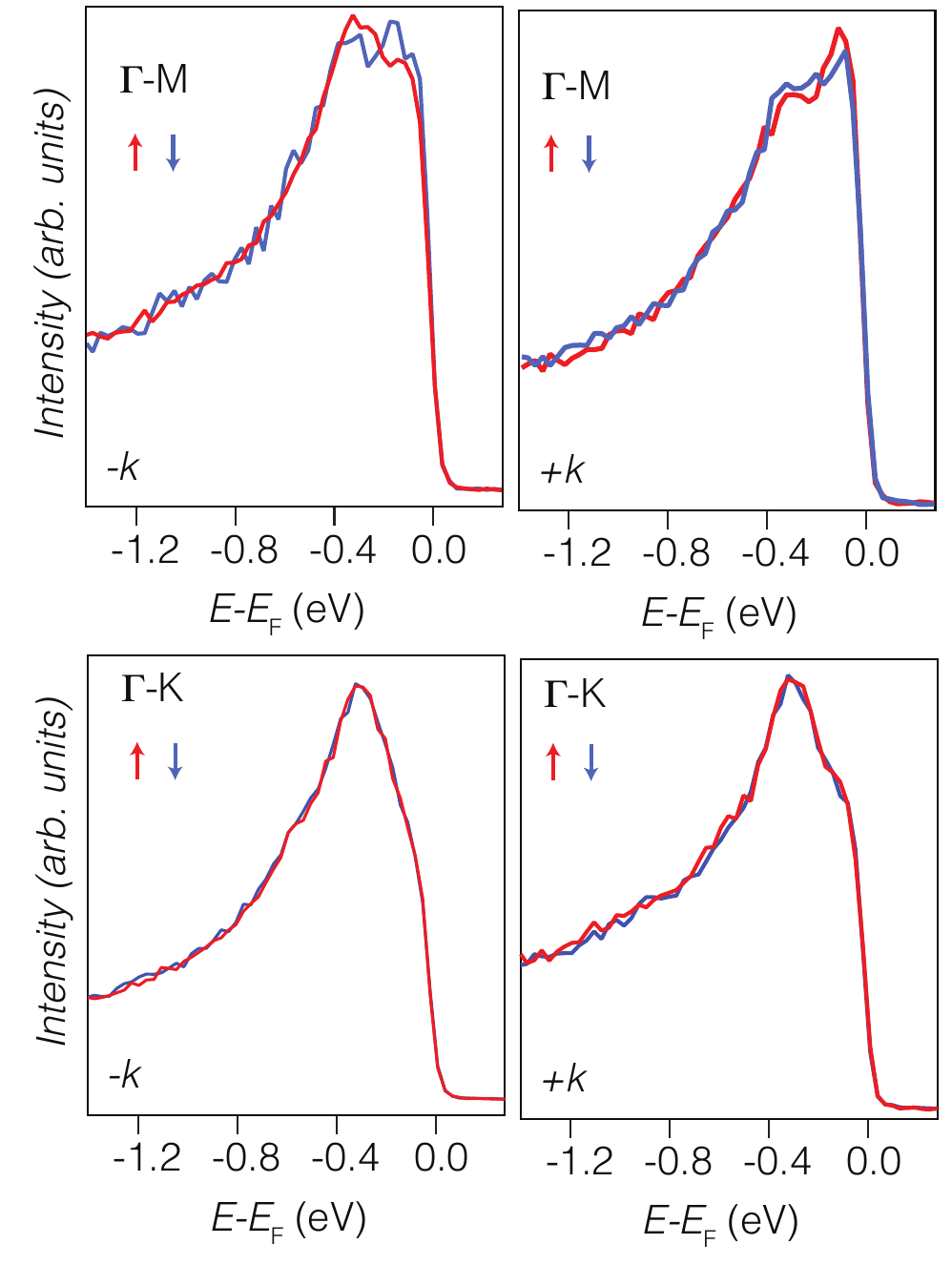}
		\caption{\textbf{Spin-ARPES along different high-symmetry directions.} Upper row shows the same results presented in the main text along the $\Gamma$-M direction, while below we show the spin-ARPES along the $\Gamma$-K, which, in agreement with the theory, gives us a zero signal.}
		\label{Domain1}
	\end{figure}
	
	To directly probe the spatial variation of domain-related spin textures, additional spin-ARPES measurements were performed at a second location on the same sample, approximately $1~\mathrm{mm}$ from the original spot. The resulting spin polarization signal exhibits an inversion of the spin-resolved spectral weight, \textit{i.e.}, the red and blue curves are reversed—consistent with switching to a domain of opposite orientation. \rev{A structural domain rotated by $60^\circ$ does not provide an alternative: through the $\{6_{001}|00\frac{1}{2}\}$ screw it is equivalent, up to a translation, to the original crystal with reversed Néel vector.} Crucially, the signal remains finite and well resolved, confirming that the domain size is large enough to produce measurable, non-zero spin polarization, which is the crucial factor for demonstrating the time-reversal connection (See Fig.\ref{Domain2}).
	
	\renewcommand{\thefigure}{S20}
	\begin{figure}[htb]
		\includegraphics[width=0.6\columnwidth]{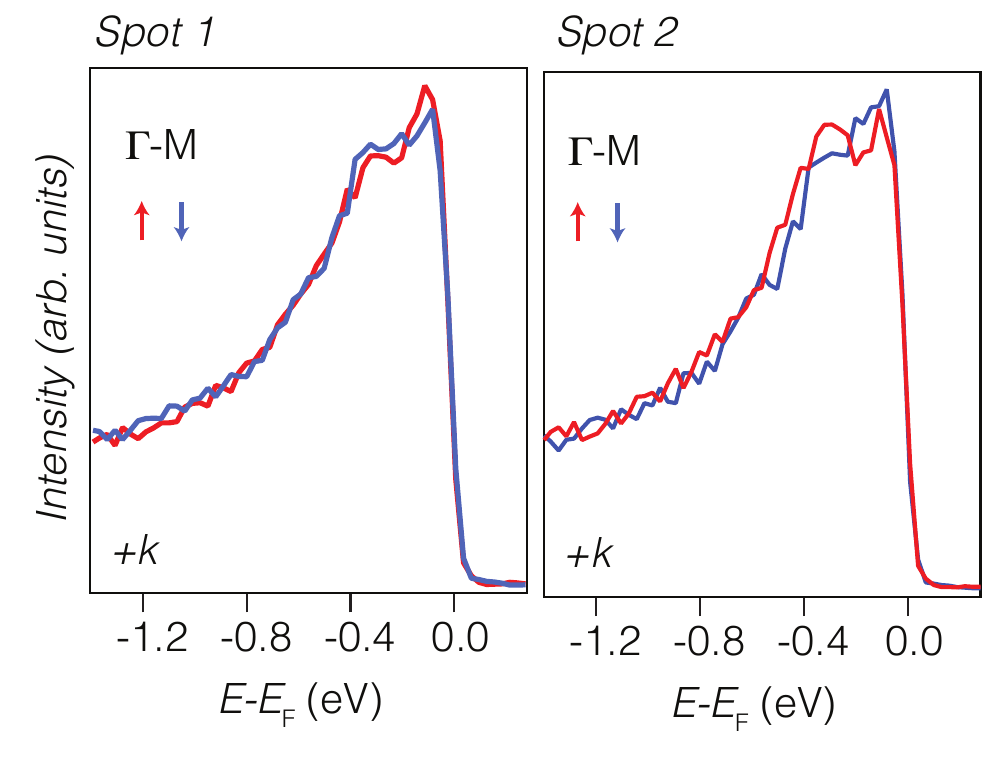}
		\caption{\textbf{Spin-ARPES across different spots.} Spin-ARPES along the $\Gamma-M$ direction performed across the same sample's surface on two different spots.}
		\label{Domain2}
	\end{figure}
	
	Spin-ARPES data were also acquired from a separately cleaved sample. These measurements again revealed a finite spin polarization, confirming the robustness and reproducibility of the observed spin texture (Fig. \ref{Domain3}).
	
	\renewcommand{\thefigure}{S21}
	\begin{figure}[htb]
		\includegraphics[width=0.6\columnwidth]{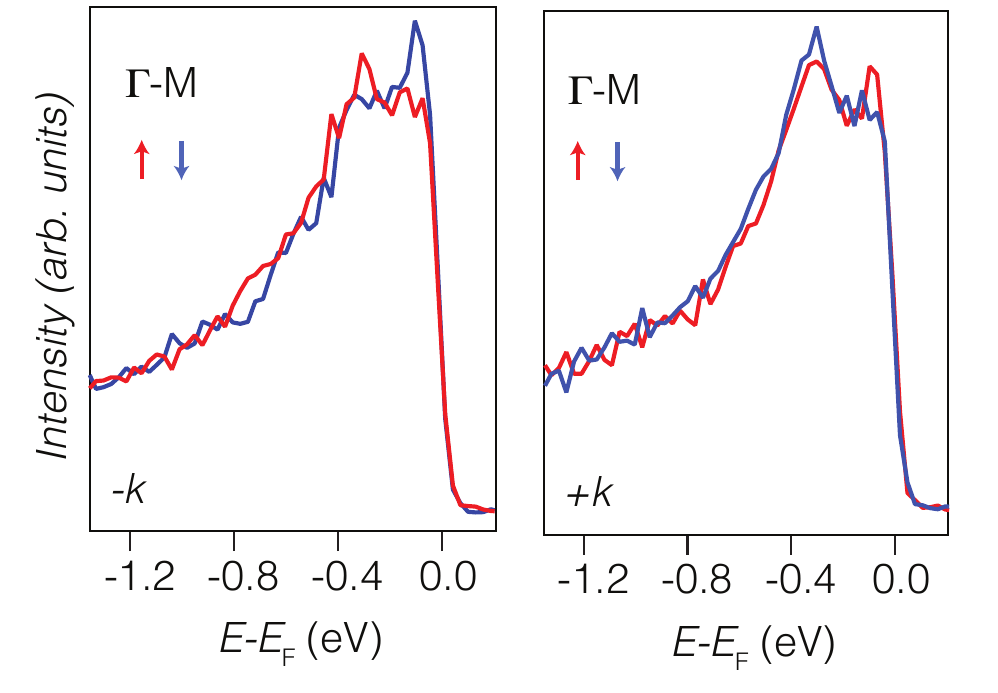}
		\caption{\textbf{Spin-ARPES measured on a different sample.} Spin-ARPES along the $\Gamma-M$ direction.}
		\label{Domain3}
	\end{figure}
	
	Taken together, these results place a statistical lower bound on the domain size, as illustrated qualitatively in Fig.~\ref{beamfootprint}.
	
	\renewcommand{\thefigure}{S22}
	\begin{figure}[htb]
		\includegraphics[width=0.6\columnwidth]{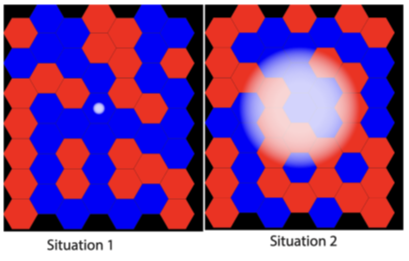}
		\caption{Situation 1: Domains are comparable to or larger than the beam size. A finite spin polarization is observed. Situation 2: Domains are significantly smaller than the beam and randomly oriented. The spin polarization averages to zero.}
		\label{beamfootprint}
	\end{figure}
	
	To demonstrate momentum-space time-reversal symmetry (TRS) breaking, it is sufficient to observe a reversal in spin polarization at time-reversal-conjugate momenta originating from the same real-space location. Since our spin-resolved spectra are collected from a single illuminated region while varying the momentum, the observed spin reversal is a direct signature of TRS breaking. In addition, observing this reversal at distinct domains further reinforces the presence of intrinsic spin-momentum locking not attributable to conventional ferromagnetic or antiferromagnetic ordering.
	
	Lastly, the statistical significance of the spin-ARPES polarization curves is shown \rev{in} Fig.~\ref{spinsi}, where the spin polarization extracted from the curves in Fig.~2i of the main text are displayed, together with $1\sigma$ statistical error bands. The errors have been computed from the counting statistics of the two V-LEED channels and propagated through the standard spin-polarisation expression after the Sherman-function correction, so that the significance of the sign reversal can be assessed directly. \rev{At $E-E_F\approx\SI{-0.05}{\eV}$ the two polarizations are $\approx+3\%$ and $\approx-9\%$; their antisymmetric part, $[P(k)-P(-k)]/2\approx6\pm1.5\%$ in magnitude, i.e.\ about four standard deviations, is the quantity required by the $g$-wave form factor, and it is insensitive by construction to any $k$-independent instrumental offset. A smaller $k$-even component ($\approx-3\%$, $\approx2\sigma$) is removed in the antisymmetric combination. Along $\Gamma$--K the polarization is compatible with zero within $1\sigma$ at both momenta (Fig.~\ref{spinsi}b).}
	
	\renewcommand{\thefigure}{S23}
	\begin{figure}[htb]
		\includegraphics[width=0.6\columnwidth]{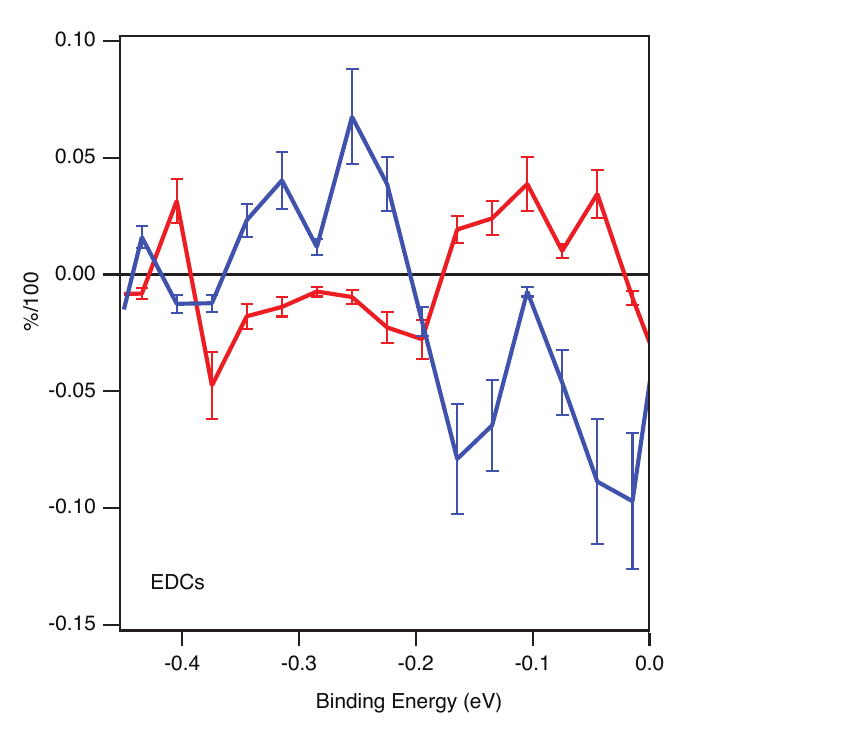}
		\caption{\rev{\textbf{a.}} Spin polarization extracted from the spin-resolved EDCs in Fig.~2i of the main text \rev{(opposite momenta along $\Gamma$--M)}, with their statistical error of 1$\sigma$. \rev{\textbf{b.} Same analysis for the spin-resolved EDCs along $\Gamma$--K of Fig.~\ref{Domain1}.}}
		\label{spinsi}
	\end{figure}

	\subsubsection*{\textbf{VIII. $k_z$ scans\rev{, photon-energy calibration, and $k_z$ broadening}}}
	
	Here, we present the $k_\parallel$ versus $k_z$ dispersion, which directly reveals a closed Fermi surface (See Fig. \ref{kz2}). This method provides an unambiguous identification of the high-symmetry points. The procedure involves plotting $k_z$ as a function of $k_\parallel$ along the $\Gamma$–M direction. In this representation, some of the bands form a closed surface, with the center identified as the $\Gamma$ point and the edges corresponding to the $A$ point. This allows for direct determination of the photon energies corresponding to the high-symmetry points. This approach has proven extremely useful, as a closed Fermi surface in $k_z$ provides the most reliable means of assessing the three-dimensional nature of the band structure and its periodicity.
	
	\renewcommand{\thefigure}{S24}
	\begin{figure}[htb]
		\includegraphics[width=0.7\textwidth]{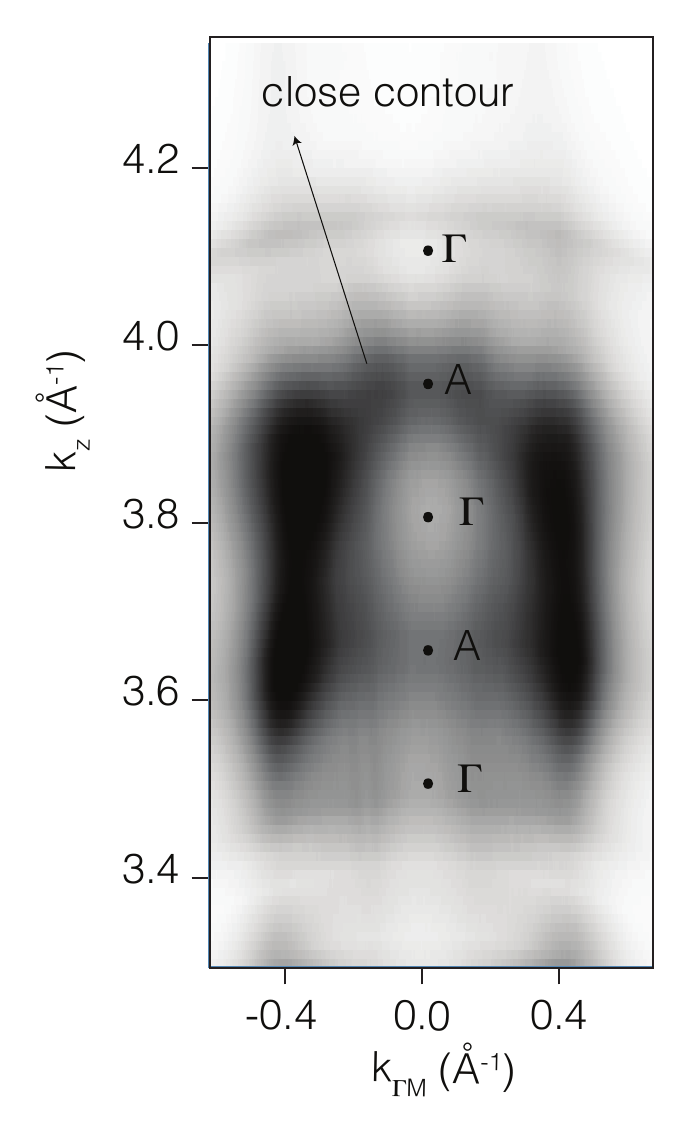}
		\caption{\textbf{Dimensionality of the electronic structure.} $k_z$ as a function of $k_\parallel$ showing the presence of a three-dimensional pocket, useful to identify the high symmetry points. \rev{Dataset: CASSIOPEE, LH polarization, $h\nu=\SIrange{25}{75}{\eV}$; $k_z$ computed with $V_0=14\pm3$~eV and $\phi=4.5\pm0.3$~eV (see text).}}
		\label{kz2}
	\end{figure}
	
	\rev{\textbf{Conversion to $k_z$.} All $k_z$ values are obtained in the free-electron final-state approximation, $k_z=0.512\,[(h\nu-\phi-E_B)\cos^2\theta+V_0]^{1/2}$ ($k_z$ in \si{\angstrom^{-1}}, energies in eV), with inner potential $V_0=14\pm3$~eV and sample work function $\phi=4.5\pm0.3$~eV. Both are properties of the sample, not of the instrument, so the same values apply to all setups (micro-ARPES at I05, high-resolution ARPES at CASSIOPEE and spin-ARPES at APE-LE). Fig.~\ref{kz2} (dataset: CASSIOPEE, LH polarization, $h\nu=\SIrange{25}{75}{\eV}$) is plotted with these parameters. The uncertainty on $V_0$ dominates and corresponds to $\approx\pm\SI{0.1}{\angstrom^{-1}}$ in the nominal $k_z$, less than half of the $\Gamma$--A distance and smaller than the intrinsic $k_z$ window discussed below. The periodicity observed in Fig.~\ref{kz2} deviates from $2\pi/c$ by $\approx5.4\%$; a deviation of this kind, larger ($\approx20\%$), is also reported for this compound in Ref.~\cite{Dale_2024}, and it is included in the uncertainty on $V_0$. With these parameters, bulk $\Gamma$ lies near 37.5 and \SI{51}{\eV}, A near 44 and \SI{75}{\eV}, and intermediate $k_z$ near 47.5 and \SI{55}{\eV}.}
	
	\rev{\textbf{$k_z$ broadening.} The finite inelastic mean free path $\lambda$ broadens the final state along $k_z$ into a Lorentzian of width $\Delta k_z=1/\lambda$ \cite{Strocov2003,Mitsuhashi2016}. The TPP-2M formula \cite{Tanuma1994} gives $\lambda\approx\SI{4.7}{\angstrom}$ at kinetic energies of 40 to \SI{70}{\eV}, hence $\Delta k_z\approx\SI{0.21}{\angstrom^{-1}}$. Because the $c$ axis is long (\SI{12.59}{\angstrom}), the whole $\Gamma$--A distance is only $\pi/c\approx\SI{0.25}{\angstrom^{-1}}$: every spectrum integrates over about 85\% of the $\Gamma$--A segment (Fig.~\ref{kzbroad}a). Photon energies are therefore associated with $k_z$ windows rather than sharp planes, and the $k_z$ assigned to a photon energy is the nominal centre of the window. Within the window the spectral weight is not uniform: photoemission matrix elements vary with $k_z$ and can enhance a portion of the window away from its centre. A spectrum nominally centred near a high-symmetry plane can thus be dominated by intermediate-$k_z$ states, and vice versa. This is the case at \SI{44}{\eV}, nominally close to A, where the split branches of Fig.~2e of the main text are nonetheless clearly resolved: the window extends to intermediate $k_z$ on both sides of A, where the matrix elements are favorable, and the spin-integrated separation of the branches does not cancel on averaging, whereas the signed spin polarization does. For this reason spin-ARPES was performed at \SI{55}{\eV}, whose window is centred at intermediate $k_z$, the same choice made in Ref.~\cite{Dale_2024}.}
	
	\rev{\textbf{Distinguishing nodal and spin-split regions.} Because the $k_z$ resolution is limited, the identification of the nodal structure does not rely on $k_z$ selectivity, but on observables that $k_z$ broadening cannot produce. (1) The vertical nodal planes containing $\Gamma$--K include the whole $k_z$ axis, so that the splitting vanishes in the bulk along $\Gamma$--K for every $k_z$ in the window, whatever its width. (2) $\Delta_s$ is odd in $k_y$ at every $k_z$: broadening reduces the amplitude of the reversal at $\pm k$ along $\Gamma$--M, but can neither create nor remove it. (3) Averaging the signed splitting $\Delta_s\propto\sin(k_zc)$, which determines the spin contrast, over the Lorentzian window centred at $k_{z0}$ gives exactly $\langle\Delta_s\rangle=\Delta_{\max}\sin(k_{z0}c)\,e^{-c/2\lambda}$ (Fig.~\ref{kzbroad}b). This is zero for windows centred on nodal planes, whatever the broadening, and reduced to $\approx26\%$ of the intrinsic value at intermediate $k_z$ ($\lambda=\SI{4.7}{\angstrom}$; $\approx35\%$ for $\lambda=\SI{6}{\angstrom}$). Together with domain averaging in the spin-ARPES spot, this accounts for the measured polarization of 5 to 10\%. (4) The reversal between the two spots of Fig.~\ref{Domain2} cannot be produced by $k_z$ broadening or by matrix elements.}
	
	\renewcommand{\thefigure}{S25}
	\begin{figure}[htb]
		\includegraphics[width=0.95\textwidth]{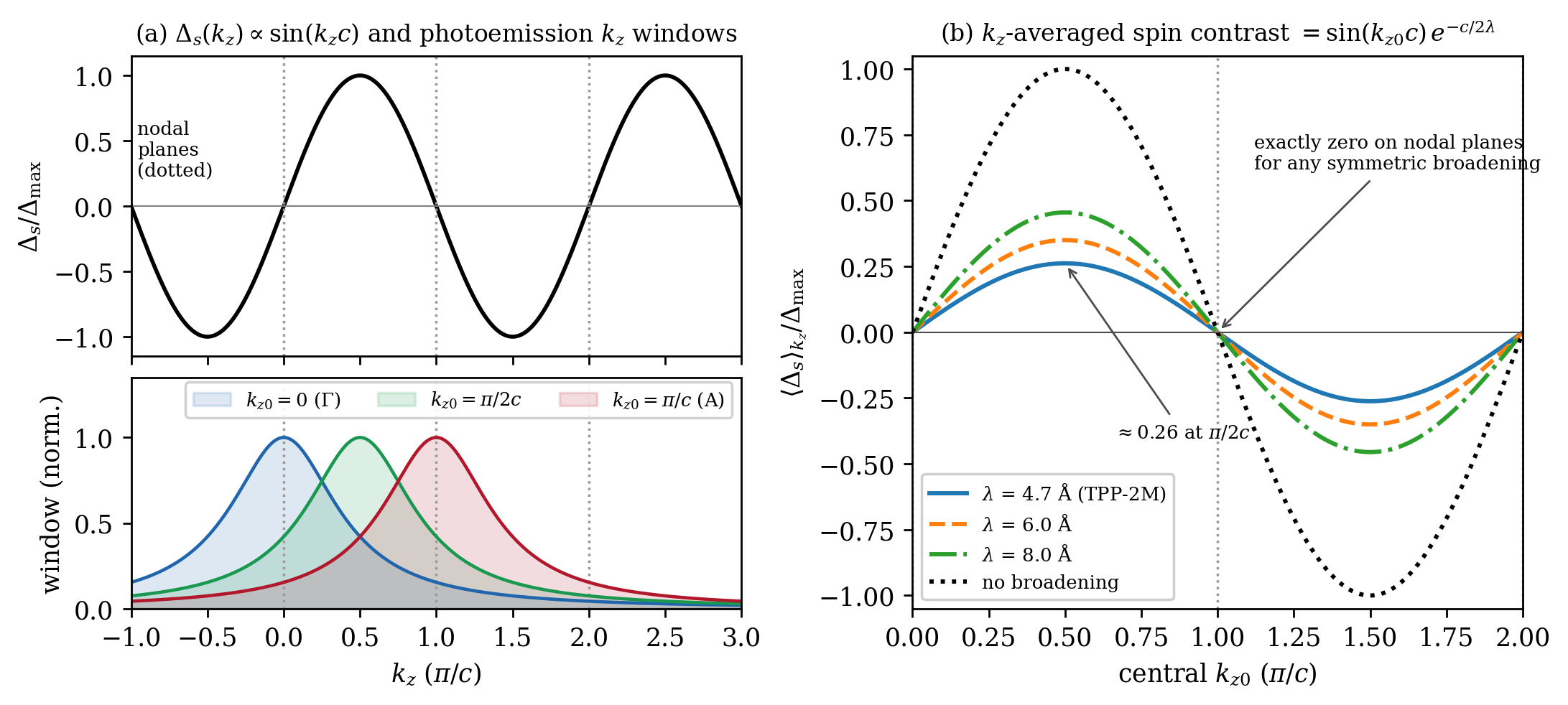}
		\caption{\rev{\textbf{Effect of the $k_z$ broadening on the calculated spin splitting.} \textbf{a.} Top: $\Delta_s\propto\sin(k_zc)$, with nodal planes at $k_z=0$, $\pi/c$ and $2\pi/c$ (dotted). Bottom: photoemission $k_z$ windows (Lorentzian of width $1/\lambda\approx\SI{0.21}{\angstrom^{-1}}\approx0.85\,\pi/c$, $\lambda=\SI{4.7}{\angstrom}$) centred at $\Gamma$, at an intermediate $k_z$ and at A. \textbf{b.} $k_z$-averaged spin contrast $\langle\Delta_s\rangle=\Delta_{\max}\sin(k_{z0}c)\,e^{-c/2\lambda}$ as a function of the window centre, for $\lambda=4.7$, 6 and \SI{8}{\angstrom} and without broadening: exactly zero for windows centred on nodal planes, with the sign preserved elsewhere.}}
		\label{kzbroad}
	\end{figure}
	
	\subsubsection*{\textbf{\rev{IX. Symmetry of relativistic and nonrelativistic $\sigma_z$ signals}}}
	
	\rev{Fig.~\ref{formfactors} compares the momentum dependence of the nonrelativistic altermagnetic splitting, $\Delta_s\propto k_y(3k_x^2-k_y^2)$ at fixed intermediate $k_z$, with that of the hidden Ising polarization of an individual NbSe$_2$ layer, which keeps $D_{3h}$ symmetry in CoNb$_4$Se$_8$ and therefore follows $k_x(k_x^2-3k_y^2)$ \cite{Zhang2014hidden,Riley2014}. The two are complementary: the former is nodal along $\Gamma$--K and finite, with opposite signs at $\pm k$, along $\Gamma$--M; the latter is nodal along $\Gamma$--M and largest along $\Gamma$--K, and is in any case expected to be strongly reduced by the covalent interlayer bonding (Section VII). More generally, any time-reversal-even $\sigma_z$ vanishes along $\Gamma$--M, because the vertical mirror containing this direction reverses $\sigma_z$ and survives at the surface. The spin-ARPES data, finite and reversing along $\Gamma$--M and compatible with zero along $\Gamma$--K, follow the altermagnetic pattern.}
	
	\renewcommand{\thefigure}{S26}
	\begin{figure}[htb]
		\includegraphics[width=0.95\textwidth]{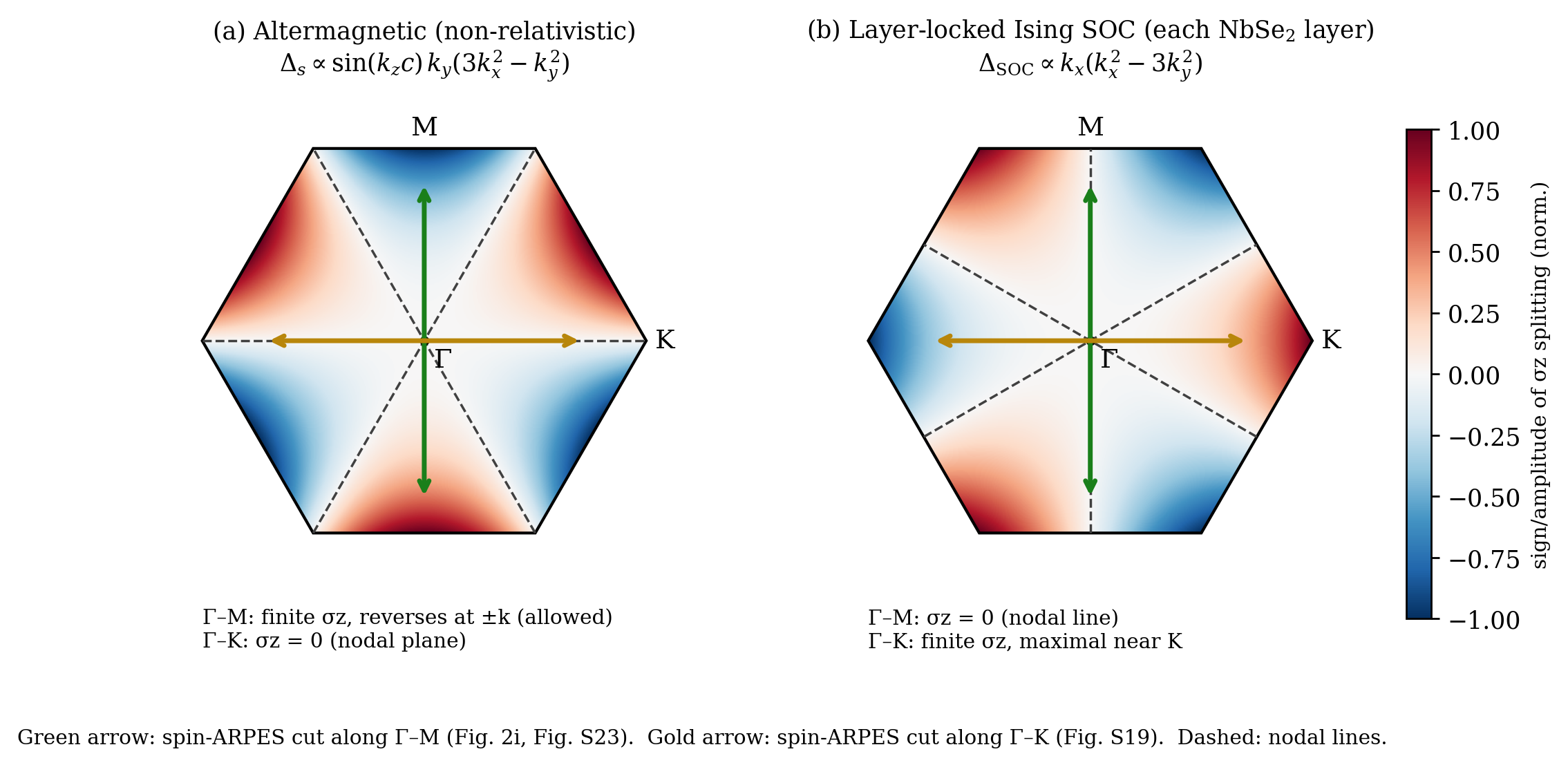}
		\caption{\rev{\textbf{Altermagnetic and SOC-induced $\sigma_z$ splitting in the $k_\parallel$ plane.} \textbf{a.} Altermagnetic form factor $k_y(3k_x^2-k_y^2)$: nodal along $\Gamma$--K, finite and reversing sign along $\Gamma$--M. \textbf{b.} Hidden Ising SOC of a NbSe$_2$ layer ($D_{3h}$), $k_x(k_x^2-3k_y^2)$: nodal along $\Gamma$--M, largest along $\Gamma$--K. Arrows: spin-ARPES cuts along $\Gamma$--M (finite, reversing signal) and $\Gamma$--K (null). Only the altermagnetic pattern matches the data.}}
		\label{formfactors}
	\end{figure}

	\newpage
	
	\bibliographystyle{MSP}
	
	\bibliography{alter.bib}